\documentclass{aa}  
\usepackage{ae,aecompl}
\usepackage{adjustbox}
\usepackage{graphicx}
\usepackage{txfonts}
\usepackage{hyperref}
\usepackage{xcolor}

\usepackage{orcidlink}

\hypersetup{colorlinks,linkcolor={blue},citecolor={blue},urlcolor={blue}}  
\usepackage{graphicx}
\usepackage{epsf,palatino,url,wrapfig,array,setspace,amsmath,amssymb,fancyhdr,multirow,lscape,appendix,rotating}
\usepackage{graphics,epsfig,txfonts}
\usepackage{threeparttable}
 
\begin{document}

\title{Black-hole mass estimation through accretion disk spectral fitting for high-redshift blazars}
   \titlerunning{Black-hole masses of high-z blazars}
   \author{G. Kyriopoulos\orcidlink{0009-0002-6889-0719}\inst{1}, M. Petropoulou\orcidlink{0000-0001-6640-0179}\inst{1,2}\thanks{\email{mpetropo@phys.uoa.gr}}, G. Vasilopoulos\orcidlink{0000-0003-3902-3915}\inst{1,2}\thanks{\email{gevas@phys.uoa.gr}}, and 
   D. Hatzidimitriou\orcidlink{0000-0002-5415-0464}\inst{1}
     }       
   \institute{Department of Physics, National and Kapodistrian University of Athens, University Campus Zografos, GR~15784, Athens, Greece
 \and
    Institute of Accelerating Systems \& Applications, University Campus Zografos, Athens, Greece
   }
   \date{Received ... ; accepted ...}

 
  \abstract
   {High-redshift ($z>2$) blazars, with relativistic jets aligned toward us, probe the most powerful end of the active galactic nuclei (AGN) population. 
   }
   {We aim at determining the black hole masses and mass accretion rates of high-$z$ blazars in a common framework that utilizes a Markov Chain Monte Carlo (MCMC) fitting method and the Shakura-Sunayev multi-temperature accretion disk model, accounting also for attenuation due to neutral hydrogen gas in the intergalactic medium (IGM).}
   {We compiled a sample of 23 high-redshift blazars from the literature with publicly available infrared-to-ultraviolet photometric data. We performed a Bayesian fit to the spectral energy distribution (SED) of the accretion disk, accounting for upper limits, and determined the black hole masses and mass accretion rates with their uncertainties. We also examined the impact of optical-ultraviolet attenuation due to gas in the IGM.}
  {We find that neglecting IGM attenuation in SED fits leads to systematically larger black-hole mass estimates and correspondingly lower Eddington ratios, with the bias becoming more severe at higher redshift. Our MCMC fits yield median black-hole masses in the range $\sim (10^{8}-10^{10})\,M_{\odot}$ and a broad distribution of median Eddington ratios ($\lambda_{\rm Edd}\sim0.04$ up to $\sim1$), with several objects accreting at $\lambda_{\rm Edd}\gtrsim0.4$ both at $z\sim2-3$ and at $z\gtrsim5$. Comparison with previous literature shows no clear method-dependent systematic offsets, although individual mass estimates can  differ by up to a factor of a few. We also demonstrate that assumptions about black-hole spin introduce a systematic degeneracy (changing the inferred $M_{\rm BH}$ and $\dot{M}$ by factors up to five each). }
   {This work is to our knowledge the first systematic study to model the accretion-disk emission of a large sample of high-$z$ blazars within a single, consistent statistical framework. Our results emphasize the importance of accounting for IGM attenuation and of using uniform fitting methods when comparing disk-based black hole estimates across samples.}
   
   \maketitle
%

\section{Introduction}
\label{sec:intro}
Blazars are active galactic nuclei (AGN) whose relativistic plasma jets are oriented at small angles to our line of sight. They display broadband, non-thermal emission extending from radio frequencies to very high-energy gamma rays. This radiation, produced by processes such as synchrotron emission and inverse Compton scattering within the jet, forms a characteristic spectral energy distribution (SED) with two broad components: a low-energy hump peaking between the infrared (IR) and X-ray frequencies, and a high-energy hump peaking in the gamma-ray regime~\citep[for a recent review, see][]{2019NewAR..8701541H}.

The central engine powering a blazar is essentially an accretion disk that channels plasma and magnetic flux to a supermassive black hole with mass $M_{\rm BH}$ in the range $10^7 M_{\odot}$ and $10^{11} M_{\odot}$. For accretion rates above a critical threshold, $\dot{m}_c \sim 0.03$ \citep{1995ApJ...452..710N}, the flow is well described by a geometrically thin, radiatively efficient disk model \citep{1973A&A....24..337S, 1973blho.conf..343N}. Here, the dimensionless accretion rate is defined as $\dot{m} \equiv \dot{M}/\dot{M}_{\rm Edd}$, where $\dot{M}_{\rm Edd}$ is the Eddington mass accretion rate, which is given by $L_{\rm Edd}/(\eta c^2) = 1.4 \times 10^{18}~(M_{\rm BH}/M_{\odot})(0.1/\eta)~\rm g \, s^{-1}$ and $\eta$ is the radiative efficiency of the accretion disk.  The resulting thermal emission is well approximated by a multi-temperature blackbody spectrum, which typically peaks in the optical–ultraviolet (OUV) range, producing the so-called Big Blue Bump; this is the most prominent component of the SED of radio-quiet AGN~\citep[e.g.][]{2013MNRAS.431..210C}.

In many blazars, this thermal component is outshone by the non-thermal jet emission, making it difficult to detect. However, it becomes visible when the synchrotron peak frequency lies below the OUV band or when the disk is intrinsically luminous, as in some low-synchrotron-peaked (LSP) blazars and powerful flat-spectrum radio quasars (FSRQs) \citep[e.g.][]{2012MNRAS.420.2899G}. High-redshift (high-$z$) blazars represent the most luminous end of the blazar population. Given the observed correlation between jet power and accretion disk luminosity~\citep[e.g.][]{2014Natur.515..376G,  2016Galax...4...36G}, blazars at $z>3$ are expected to host more luminous accretion disks and black holes with $M_{\rm BH} \gtrsim 10^9 M_{\odot}$ \citep[e.g.][]{2011MNRAS.416..216V,  2025ApJ...985..179S, 2025SSRv..221...62S}. Accurate measurements of black hole masses in such systems are essential for constraining the formation and growth channels of supermassive black holes \citep[for a review, see][]{2016PASA...33...54R}.
 
Previous studies have often estimated disk properties either by visual (`by eye') fits to the accretion disk SED \citep[e.g.][]{2013MNRAS.428.1449G}, or by modeling the entire blazar SED \citep[e.g.][]{2010MNRAS.405..387G, 2020ApJ...889..164M}  for a handful of sources. Alternatively, spectroscopic virial mass estimates for blazars based on broad emission lines were derived for individual sources \citep[e.g.,][]{2020ApJ...889..164M, Belladitta2022}. 
These virial methods rely on measuring the width of emission lines originating in the broad line region (BLR), under the assumption that the BLR clouds are gravitationally bound to the central black hole~\citep[see e.g.][for such estimates in quasars]{2011ApJS..194...45S}. In this work, we present estimates of black hole mass and mass accretion rate for a sample of previously studied high-$z$ blazars within a single homogeneous framework. Using publicly available infrared-to-ultraviolet photometric data, we adopt a Bayesian SED-fitting method that accounts for upper limits and the attenuation of ultraviolet photons by intervening neutral hydrogen gas in the intergalactic medium (IGM). This statistically robust approach allows us to investigate parameter degeneracies, particularly between black hole mass and accretion rate, and to assess systematic biases in black hole mass estimates arising from IGM attenuation.

This paper is structured as follows. We outline the blazar sample in Sect.~\ref{sec:sample} and describe the model and fitting procedure in Sect.~\ref{sec:methods}. We present the results in Sect.~\ref{sec:results}, we discuss certain aspects of our analysis in Sect.~\ref{sec:discussion}, and conclude with Sect.~\ref{sec:conclusions}. Luminosity distances and black hole growth times are calculated in the framework of a flat cosmology with $H_0 = 67.8$~km s$^{-1}$~Mpc$^{-1}$ and $\Omega_{\rm M} = 0.308$ \citep{Planck2016}.
 
\section{The sample}
\label{sec:sample}

\begin{table*}
\centering
\caption{High-$z$ blazars and blazar candidates with bibliographical information.}
\label{tab:sample}
\resizebox{\textwidth}{!}{%
\begin{tabular}{c l ccc cccc cc}
\hline
No. & Name & RA (J2000) & Dec (J2000) & $z$ & $M_{\mathrm{BH}}$ ($M_{\odot}$) & $L_{\mathrm{disk}}$ (erg/s) & $\dot{M}/\dot{M}_{\mathrm{Edd}}$ & Method & Ref. \\
\hline
1 & PSO J0309+27 & 10:26:23.62 & +25:42:59.43 & 6.10 & $1.45^{+1.89}_{-0.85} \times 10^{9}$ & $8 \times 10^{46}$ & 0.22\tablefootmark{a}
 & spec & (1) \\
2 & B2 1023+25 & 10:26:23.62 & +25:42:59.43 & 5.30 & $2.8 \times 10^{9}$ & $9 \times 10^{46}$  & 0.25 & sed & (2) \\
3 & SDSS J0131-0321 & 01:31:27.31  & -03:20:59.63 & 5.18 & $1.1 \times 10^{10}$ & $4.2 \times 10^{47}$ & 0.25 & sed & (3)  \\
&  & & &  &  $2.7^{-0.4}_{+0.5}\times 10^9$ & $1.1^{+0.2}_{-0.2} \times 10^{48}$ &  3.14 & spec & (4) \\
4 & GB6 B1428+4217 & 14:30:23.74 & +42:04:36.49 & 4.715 & $1.7\times10^9$ & $2\times 10^{47}$ & - & sed & (5) \\
& & & & & $7.4\times 10^9$ & $2\times 10^{47}$ & 0.9 & spec & (10) \\ 
& & & & & $1.7\times 10^9$ & $2\times 10^{47}$ &  0.2 & spec & (10) \\ 
5& PMN J2134-0419 & 21:34:12.01  & -04:19:09.87 & 4.35 & $1.8 \times 10^{9}$ & $9.5 \times 10^{46}$ & 0.35 & sed & (6) \\
6& J151002+570243 & 15:10:2.92  & +57:02:43.38 & 4.31 & $6.5 \times 10^9$ & $7.1\times 10^{46}$ & 0.09 & sed & (7) \\
& & & & & $1.5 \times 10^{10}$ & $4\times 10^{47}$ & 0.21 & sed & (8) \\
& & & & & $3\times10^8$ & - & - & spec & (9) \\
& & & & & $3\times10^9$ & $4.5\times10^{46}$ & $0.12$ & sed & (9) \\
& & & & & $8.8 \times 10^8$ & $1.5\times 10^{47}$ & 1.3 & spec & (10) \\
7& J222032.50+002537.5 & 22:20:32.50  &  +00:25:37.51 & 4.22 & $2 \times 10^{9}$ & $4.5 \times 10^{46}$ &  0.15 & sed & (6) \\
8& J142048.01+120545.9 & 14:20:48.01 & +12:05:45.98 & 4.03 & $2 \times 10^{9}$ & $5.3 \times 10^{46}$ &  0.18 & sed & (6) \\
9& J135406-020603 & 13:54:6.90 & -02:06:4.00 & 3.71 & $10^9$ & $5.5\times 10^{46}$ & 0.43 & sed & (7) \\
& & & & & $8.9\times 10^8$ & - & - & spec & (9) \\
10 & J020346+113445 & 02:03:46.70  & +11:34:44.00 & 3.63 & $3 \times 10^9$ & $5.5\times 10^{46}$ & 0.14 & sed & (7) \\
11& J013126-100931 & 01:31:26.71 &-10:09:31.17 & 3.51 & $6 \times 10^9$ &  $5.0\times 10^{46}$ & 0.06 & sed & (7) \\
12 &J064632+445116 & 06:46:32.03  &+44:51:16.59 & 3.41 & $1.5 \times 10^9$ & $1.4\times 10^{47}$ & 0.74 & sed & (7) \\
& & & & & $1.3\times 10^9$ & - & - & spec & (9)\\ 
& & & & & $4\times 10^9$ & $1.3\times 10^{47}$ & $0.26$ &  sed & (9)\\ 
13 &S5 0014+813 & 00:17:08.47& +81:35:08.13 & 3.366 & $4 \times 10^{10}$ & $3 \times 10^{48}$ &  0.5 & sed & (11) \\
14 & J212912-153841 & 21:29:12.10 &-15:38:42.00 & 3.26 & $5 \times 10^9$ & $5.1\times 10^{47}$ & 0.79 & sed & (7) \\ 
& & & & & $10^{10}$ & $7.5 \times 10^{47}$ &  0.5 & sed & (11) \\
& & & & & $13.4^{+2.4}_{-2.4}\times 10^9$ & $1.8\times 10^{48}$ & 1.05  & spec & (12) \\
& & & & & $6.3\times 10^9$ & - & - & spec & (9) \\
& & & & & $7\times 10^9$ & $6\times 10^{47}$ & 0.68 & sed & (9) \\
15 & PKS 0537-286 & 05:39:54.28 & -28:39:55.95 & 3.104 & $2 \times 10^{9}$ & $5.4 \times 10^{46}$ &  0.18 & sed & (11) \\
16 & GB6 0805+614 & 08:05:18.18 &+61:44:23.70 & 3.033 & $1.5 \times 10^{9}$ & $3.4 \times 10^{46}$ & 0.15 & sed & (11) \\
17 & PKS 0519+01 & 05:22:17.47 & +01:13:31.18 & 2.94 & $4.5 \times 10^{9}$ & $2.36 \times 10^{46}$ &  0.035 & sed & (13) \\
18 & PMN J1344-1723 & 13:44:14.40  & -17:23:40.39 & 2.49 & $1.5 \times 10^{9}$ & $1.23 \times 10^{46}$ &  0.055 & sed & (13)  \\
19 & TXS 1149-084 & 11:52:17.21  &-08:41:03.31 & 2.37 & $1.5 \times 10^{9}$ & $3.15 \times 10^{46}$ &  0.14 & sed & (13) \\
20 & PKS 2149-306 & 21:51:55.52& -30:27:53.70 & 2.345 & $5 \times 10^{9}$ & $1.5 \times 10^{47}$ & 0.2 & sed & (11) \\
21 & 4C 71.07 & 08:41:24.37 & +70:53:42.17 & 2.172 & $3\times 10^{9}$ & $2.25 \times 10^{47}$ &  0.5 & sed & (11) \\
22 & PKS 0227-369 & 02:29:28.45& -36:43:56.82 & 2.12 & $2 \times 10^{9}$ & $3 \times 10^{46}$ &  0.1 & sed & (14) \\
23 & PKS 0528+134 & 05:30:56.42 & +13:31:55.15 & 2.04 & $10^{9}$ & $7.5 \times 10^{46}$ &  0.5 & sed & (14) \\
\hline
\end{tabular}
}
\tablefoot{
\tablefoottext{a}{\citet{Belladitta2022} report an Eddington ratio of $\sim$0.44 when using the bolometric luminosity. Since they estimate the disk luminosity to be approximately half of $L_{\mathrm{bol}}$, the corresponding Eddington ratio based on $L_{\mathrm{disk}}$ is $\sim$0.22, which we report here. }
}
\tablebib{
(1)~\citet{Belladitta2022}, (2)~\citet{2013MNRAS.433.2182S}, (3) ~\citet{2015MNRAS.450L..34G}, (4)~\citet{2014ApJ...795L..29Y}, (5)~\citet{2025ApJ...990..206G}, (6)~\citet{2015MNRAS.446.2483S}, (7)~\citet{2020ApJ...889..164M}, (8)~\citet{2024ApJ...974...38G}, (9)~\citet{2017ApJ...837L...5A}, (10)~\citet{2022MNRAS.511.5436D}, (11)~\citet{2010MNRAS.405..387G}, (12)~\citet{2004ApJ...611..761D}, (13)~\citet{2013MNRAS.428.1449G}, (14)~\citet{2011MNRAS.411..901G}.
}
\end{table*} 

Our sample consists of high-$z$ blazars drawn from previous works, including \cite{2010MNRAS.405..387G,2011MNRAS.411..901G,2013MNRAS.428.1449G, 2015MNRAS.450L..34G}, \cite{2013MNRAS.433.2182S,2015MNRAS.446.2483S}, \cite{2020ApJ...889..164M}, \cite{Belladitta2022}, and \cite{2025ApJ...990..206G}. 
It comprises 23 blazars spanning the redshift range \(2.0 \lesssim z \lesssim 6.1\). It includes eight sources at \(z > 4\) and three at \(z > 5\), representing some of the most distant and luminous blazars currently known. The sample with bibliographical information is presented in Table~\ref{tab:sample}.

For this work we compiled archival infrared-to-ultraviolet photometry primarily via the ASI/SSDC SED Builder tool \citep{Stratta2011}.  Photometric points were drawn from public surveys and catalogs, such as WISE \citep{2010AJ....140.1868W}. Additional photometry from optical/near-IR catalogs (e.g., SDSS, 2MASS, USNO A2) and database compilations (e.g., NED) were included when available. Furthermore, OUV data from the UltraViolet and Optical Telescope \citep[UVOT,][]{2005SSRv..120...95R} on board Swift were incorporated into the SED Builder tool \citep{2012A&A...541A.160G}, when present. Infrared and OUV data from all external catalogs provided by the ASI/SSDC SED Builder are corrected for Galactic extinction using the extinction law of \cite{1989ApJ...345..245C} and the H\,\textsc{i} column density from the HI4PI Map \citep{2016A&A...594A.116H}. For the subset of sources studied by \cite{2020ApJ...889..164M}, we adopted their compiled multiwavelength data set and incorporated it into our MCMC fits.  We note that publicly available OUV data of high-$z$ blazars retrieved from the SED Builder tool have not been corrected for IGM gas attenuation.

We initially used the SED Builder tool and selected photometric measurements from available catalogs with a $3''$ tolerance. We visually inspected the sky images at multiple filters and searched for wrong associations and/or contamination from nearby bright sources. More details are provided in Appendix \ref{app:phot_examples}.
A complete record of the specific catalogs and databases from which photometric points were drawn, together with the corresponding source-finding charts in multiple filters (used to verify associations), is provided in a Zenodo repository \footnote{\url{https://zenodo.org/records/17781963}}.

\section{Methods} \label{sec:methods}
In this section we summarize the methods used in previous studies of high-$z$ blazars (see Table~\ref{tab:sample}), and then we outline the SED model and the fitting procedure applied to the archival data in this work.

\subsection{Previous works} \label{sec:previous}
The methods originally employed to infer \(M_{\rm BH}\) vary across the literature. In \cite{2010MNRAS.405..387G,2011MNRAS.411..901G,2015MNRAS.450L..34G}, the approach consisted of full broadband SED modeling, simultaneously fitting the OUV accretion-disk emission with a Shakura–Sunyaev model and the X-ray/$\gamma$-ray emission with a one-zone leptonic synchrotron self-Compton plus external Compton (SSC+EC) jet model; in those works the model parameters were adjusted to reproduce the observed SEDs (parameter tuning and visual, “by-eye” comparison), allowing for the derivation of the black hole mass and accretion rate. \citet{2010MNRAS.405..387G} used an effective IGM opacity, $\tau_{\rm eff}(z)$, by averaging the transmitted flux over an ensemble of lines of sight and using the gas distribution of \cite{2012ApJ...746..125H}. They did so to correct rest-frame UV photometry for attenuation by intervening H\,\textsc{i} (Lyman-series and Lyman-continuum) systems at high redshift, which otherwise suppresses the observed disk peak and can bias estimates of $L_{\rm disk}$ and $M_{\rm BH}$ derived from disk SED fitting.

\cite{2013MNRAS.428.1449G} obtained quasi-simultaneous observations for five Fermi-detected blazars at \(z>2\) with the Gamma-Ray burst Optical Near-Infrared Detector (GROND) instrument \citep{2008PASP..120..405G}, UVOT, and the X-Ray Telescope \citep[XRT,][]{2005SSRv..120..165B} on board the Neil Gehrels Swift satellite. They detected a thermal accretion-disk component in four cases, and derived \(M_{\rm BH}\) and \(\dot{M}\) from a Shakura–Sunyaev fit to the near-IR/UV bump (without reporting uncertainties to the fitted parameters). 

In the absence of such a disk signature, they estimated \(L_{\rm disk}\) either from the BLR luminosity $L_{\rm BLR}$ (assuming \(L_{\rm disk}\simeq 10\,L_{\rm BLR}\)) or from monochromatic continuum measurements such as \(\lambda L_{\lambda}\) at 3000\,\AA\ with an empirical bolometric correction; this was the case for PKS\,0227$-$369 (\(z=2.12\)) and PKS\,0528+134 (\(z=2.04\)), where no disk bump was visible. \cite{2015MNRAS.446.2483S} likewise fitted a Shakura–Sunyaev disk directly to the optical–infrared SED when the blue bump was apparent, as in PMN\,J2134$-$0419. \cite{2020ApJ...889..164M} performed broadband blazar SED modeling for all their sources, jointly treating the accretion disk and jet components to ensure uniform parameter estimation. In the latter studies, no uncertainties on the fitted parameters were reported. \cite{Belladitta2022} applied two independent techniques to PSO\,J0309+27 (\(z \approx 6.1\)): a single-epoch virial estimate based on the C\,\textsc{iv} \(\lambda1549\) line, and an accretion disk fit constrained by near-infrared photometry from Telescopio Nazionale Galileo (TNG) and Pan-STARRS1 (PS1), finding agreement between the two determinations.

For completeness, we note that all other spectroscopic measurements listed in Table~\ref{tab:sample} are derived using the virial method applied to the BLR clouds.  In this approach, the black hole mass is inferred from the velocity dispersion of the BLR clouds and the size of the BLR, $R_{\rm BLR}$. The latter can be obtained by a single-epoch measurement of either the continuum or line luminosity, $L$, using scaling relations of the form $R_{\rm BLR}\propto L^a$ \citep[e.g.,][]{2000ApJ...533..631K}. The velocity dispersion is determined from the width of broad emission lines, most commonly the C\textsc{iv}~\(\lambda1549\) line, as it lies within the optical spectral range for quasars up to $z\sim 5.5$ \citep{2022MNRAS.511.5436D}. For a detailed description of the spectroscopic measurements and adopted calibrations, we refer the reader to the respective works.

\subsection{SED Model}\label{sec:model}
We adopt the geometrically thin and optically thick accretion disk model of \cite{1973A&A....24..337S}. According to this, each annulus of the disk at radius $r$ emits as a blackbody of temperature,
\begin{equation}
T(r) = \left[ \frac{3 \dot{M} R_{\rm g} c^2}{8 \pi \sigma r^3}\left(1 - \sqrt{\frac{R_{\rm in}}{r}} \right) \right]^{1/4}
\label{eq:Tr}
\end{equation}
where $\sigma$ is the Stefan-Boltzmann constant, $\dot{M}$ is the mass accretion rate, $R_{\rm g} = GM_{\rm BH}/c^2$ is the gravitational radius of a black hole with mass $M_{\rm BH}$, and $R_{\rm in}= 6 R_{\rm g}$ is the inner disk radius (in Sect.~\ref{sec:spin} we discuss how a different inner disk radius, due to a non-zero black hole spin, would affect our results). The differential flux $F_{\nu}$ of the accretion disk received by an observer at a viewing angle $\theta$ (angle between the line of sight and normal to the disk) is given by the superposition of blackbody spectra with varying temperature~\citep{2002apa..book.....F}:
\begin{equation}
 F^{AD}_\nu =\frac{ 2 \pi \cos\theta }{d_L^2} \int_{R_{\text{in}}}^{R_{\text{out}}} \!\!\! {\rm d}r \, r  \, B_{\nu}(T(r)),
 \label{eq:Fdisk}
\end{equation}
where $d_L$ is the luminosity distance of the source, $B_\nu$ is the Planck function, $R_{\rm in} = 6 R_{\rm g}$ and $R_{\rm out} = 2000\, R_{\rm g}$ are respectively the inner and outer disk radii. Because we are modeling blazars, which by definition are viewed at small angles, we fixed $\theta$ to 3 degrees ($\cos \theta \approx 1$) without loss of generality. 

For sources where the non-thermal emission of the jet is present in the infrared-to-optical part of the spectrum (after visual inspection), we consider an additional empirical component.  This is described as a power law with slope $-s$ followed by an exponential cutoff at frequency $\nu_{c}$, mimicking the synchrotron spectrum emitted by a power-law distribution of particles,

\begin{equation}
F^{jet}_{\nu} = F_0 \, \nu_0^{s-1} \nu^{-s} e^{-\frac{\nu}{\nu_c}}, 
\label{eq:Fjet}
\end{equation}
where the pivot frequency is $\, \nu_0 = 10^{13.7}~\rm Hz$, and $F_0$ is the normalization of the spectrum in $\nu F_\nu$ units.

\begin{figure}
    \centering
    \includegraphics[width=0.4\textwidth]{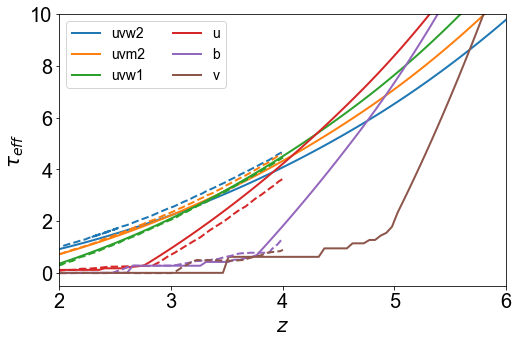}
    \caption{Effective opacity (averaged over an ensemble of line of sights)  due to neutral hydrogen gas in the intergalactic medium as a function of redshift for different wavelengths corresponding to the six Swift-UVOT filters. Results obtained from the analytical model of \cite{2014MNRAS.442.1805I} are plotted with solid lines. Results adopted from Figure 3 of \cite{2010MNRAS.405..387G} are overplotted for comparison with dashed lines.}
    \label{fig:tau_eff}
\end{figure}

Observations of high-$z$ blazars in OUV light may be affected by attenuation from neutral hydrogen left in the IGM \citep[e.g.][]{1965ApJ...142.1633G}. For example, photons observed with Swift UVOT would be susceptible to attenuation by neutral hydrogen for sources at $z>2$. Publicly available OUV observations are usually not corrected for the IGM attenuation. As a result, we have to account for this source of attenuation in the fitting method. An accurate calculation of the opacity requires knowledge of the spatial distribution of the gas density in the IGM along a specific line of sight. \cite{2010MNRAS.405..387G} have calculated an effective opacity $\tau_{\rm eff}$ by averaging over all possible lines of sights, following the gas density distribution of \cite{2012ApJ...746..125H}. Here, we adopt the analytical model of \cite{2014ascl.soft02019I, 2014MNRAS.442.1805I} that provides $\tau_{\rm eff}(\lambda_{\rm obs}, z)$ and can be easily incorporated into our fitting procedure. A comparison of the effective opacity adopted by \cite{2010MNRAS.405..387G} and the one inferred by the model of \cite{2014MNRAS.442.1805I} is presented in Fig.~\ref{fig:tau_eff}.  Differences, which are driven mainly by the different underlying gas distribution, are more evident in the $\rm b,v$ UVOT filters, where the absorption edges are more prominent. However, in ultraviolet filters, where the opacities are large even at $z=2$, the differences are negligible.

In summary, the model that will be fitted to the photometric data is written as $\nu F_\nu = (\nu F^{AD}_{\nu} + \nu F^{jet}_{\nu})\, e^{-\tau_{\rm eff}(\nu, z)}$ or without the jet contribution if there is no evidence of a soft power law in the data. 

\subsection{Fitting procedure}\label{sec:fit}
Our goal is to derive the posterior probability density $p$ from the Bayes’ theorem  given a data set ($D$) and priors for a model with parameters contained in the vector $\vartheta$ \citep{2005blda.book.....G, 2010arXiv1008.4686H}. The likelihood function reads:

\begin{equation}
\ln{\mathcal{L}({\vartheta}|D)} =  -\frac{1}{2} \sum_{i}^{} \left[\frac{(\phi_{\rm m,i}-\phi_{\rm d, i})^2}{\sigma_{\rm tot, i}^2} + \ln{\sigma_{\rm tot, i}^2} \right], 
\end{equation}

where $i$ runs over the number of observations, $\phi \equiv \log_{10}(\nu F_{\nu})$, and the subscripts $m, d$ refer to the model expectation and data, respectively. The comparison of the model and data is performed in the rest frame of the blazar\footnote{The emitted photon frequency $\nu_{\rm em}$ is related to the observed one as $\nu_{\rm em} = \nu (1+z)$, and $\nu F_{\nu} = \nu_{\rm em} F_{\nu_{\rm em}}$.}, and $\sigma_{\rm tot}$ is defined as

\begin{equation}
\label{eq:sigma_tot}
\sigma_{\rm tot, \rm i}^2 = \sigma_{\rm i}^2 + e^{2\ln{f}}.
\end{equation}

Here,  $\sigma_{\rm i}$ is the statistical uncertainty of the flux measurement $i$ (in logarithm). We also introduced the term $\ln f$ to account for other sources of uncertainty that are not captured by $\sigma_{\rm i}$. For example, 
scatter in the photometric data may arise from long-term variability of the disk and/or jet. Moreover, photometric fluxes represent averages over the filter bandwidth, while model fluxes are evaluated at a single frequency corresponding to the filter’s central wavelength. In addition, the use of the term $\ln f$ avoids pathological cases where no statistical errors are reported for the archival photometric data.  

Upper limits on fluxes are also included in the fit by adding the following term to the likelihood function \citep[see][]{2012PASP..124.1208S,2020MNRAS.491..740Y}:
\begin{equation}
\sum_j^{} \ln \left ( \sqrt{\frac{\pi}{2}}\sigma_{\rm j} \left [1+\textrm{erf}\left(\frac{\phi_{\rm lim,j}-\phi_{\rm m,j}}{\sqrt{2} \sigma_{\rm j}}\right) \right] \right),
\end{equation}
where $\phi_{\rm lim, j}$ is the $j$-th flux upper limit (in logarithm) and $\sigma_{\rm j}$ is the respective uncertainty. 

Finally, the vector summarizing the parameters of the two-component model is $\vartheta = \{ M_{\rm BH}, \dot{M}, F_0, s, \nu_{c}, \ln f \}$, which reduces to $ \{ M_{\rm BH}, \dot{M}, \ln f \}$ for the single accretion disk model.  

To estimate the physical parameters of the accretion disk, we performed a Markov Chain Monte Carlo (MCMC) sampling using the {\tt emcee} algorithm \citep{2013PASP..125..306F}. 
We used uniform priors in logarithmic space for all parameters except the power-law slope (which was uniform in linear space). The mass accretion rate was limited to the Eddington value, since the adopted disk model is not appropriate for describing super-Eddington accretion; see Table~\ref{tab:priors} for the adopted ranges. We used 20 walkers, 10,000 (5,000) steps for the two- (one-)component models, and discarded an appropriate number of points (typically 500 to 1000) in the burn-in phase\footnote{For each chain of length $N$, we also computed the integrated autocorrelation time $\tau$ and found $N>(40-50) \tau$,  which is generally taken as evidence of convergence.}. This ensures that we only keep walkers from the converged solution.

\begin{table}
    \caption{Uniform priors used in the MCMC fit.}
    \label{tab:priors}
    \centering
    \begin{tabular}{l l}
    \hline
     Parameter & Range  \\
     \hline 
     $\log_{10} M_{\rm BH}~(M_\odot)$    & $\mathcal{U}$[8, 11] \\ 
     $\log_{10} \dot{M}~(M_\odot/yr)$    & $\mathcal{U}$[0, $\log_{10}(\dot{M}_{\rm Edd})$] \\ 
     $\log_{10}F_0$~(cgs units) & $\mathcal{U}$[-13, -10] \\ 
     $\log_{10} \nu_c$~(Hz) & $\mathcal{U}$[13.5, 14.5] \\     
     $s$ & $\mathcal{U}$[0.5, 2.5] \\ 
     $\ln f$ & $\mathcal{U}$[-6, -0.5] \\ 
     \hline
    \end{tabular}
\end{table}

\begin{figure*}
    \centering
    \includegraphics[width=0.33\textwidth]{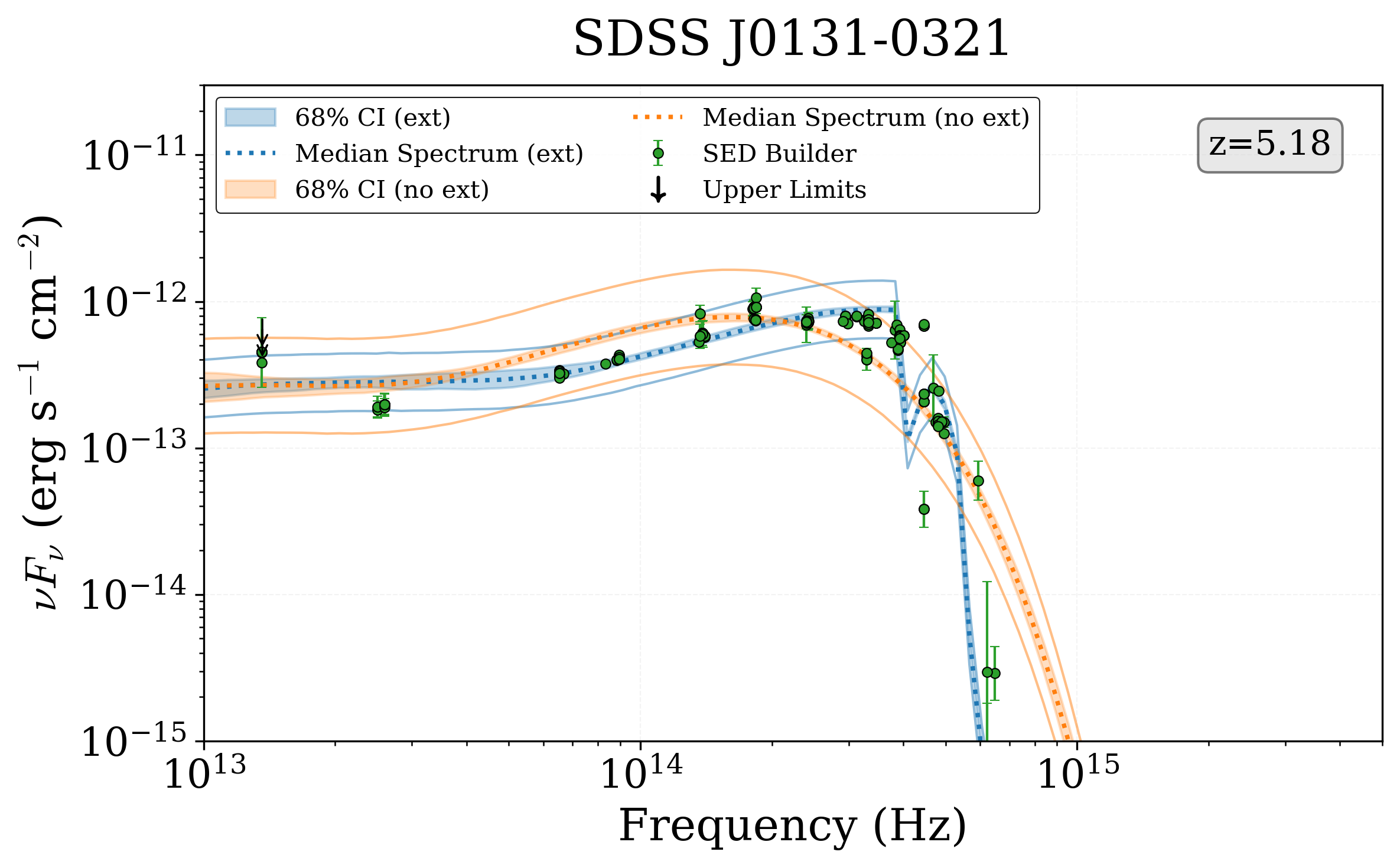}
    \includegraphics[width=0.33\textwidth]{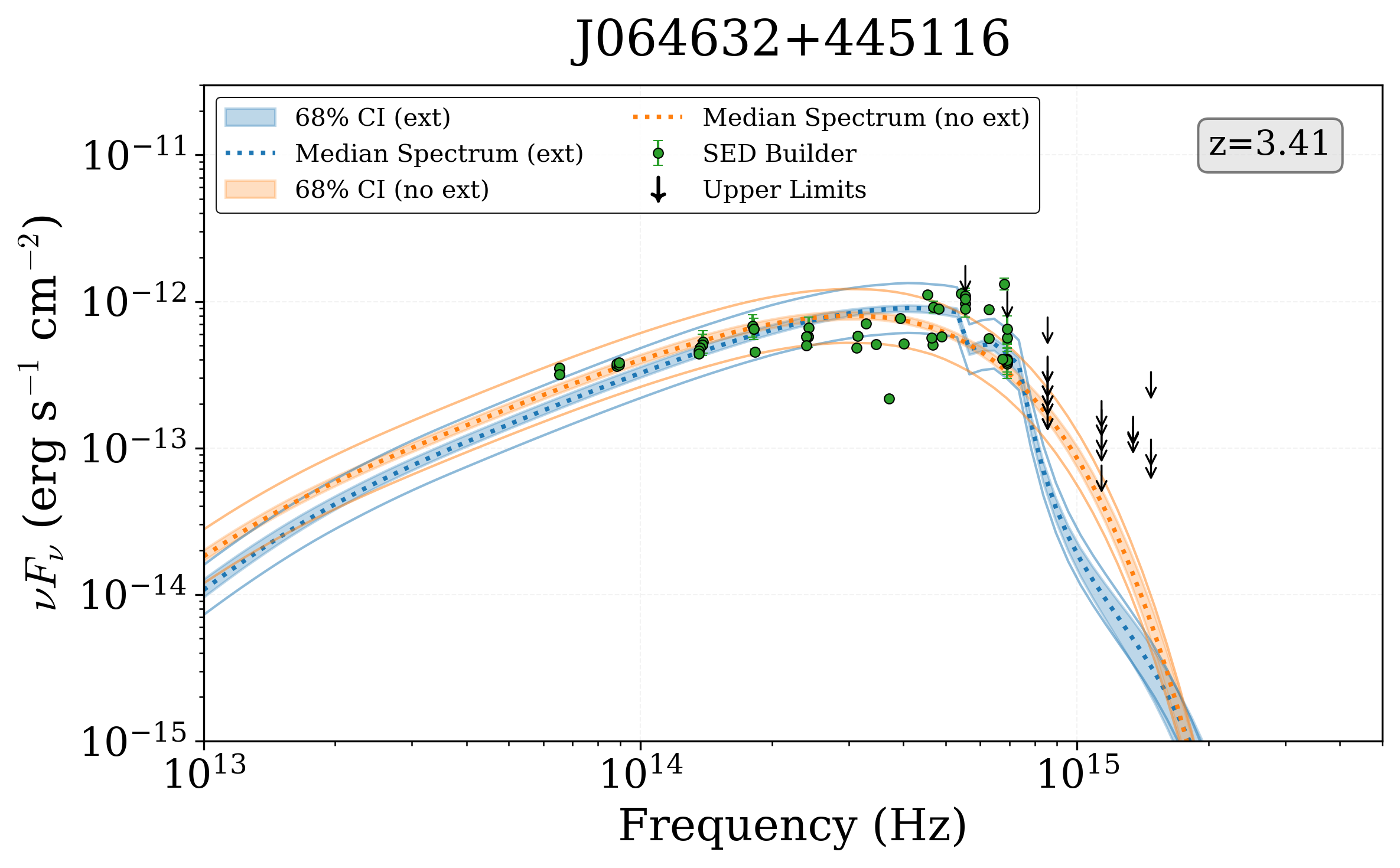}
    \includegraphics[width=0.33\textwidth]{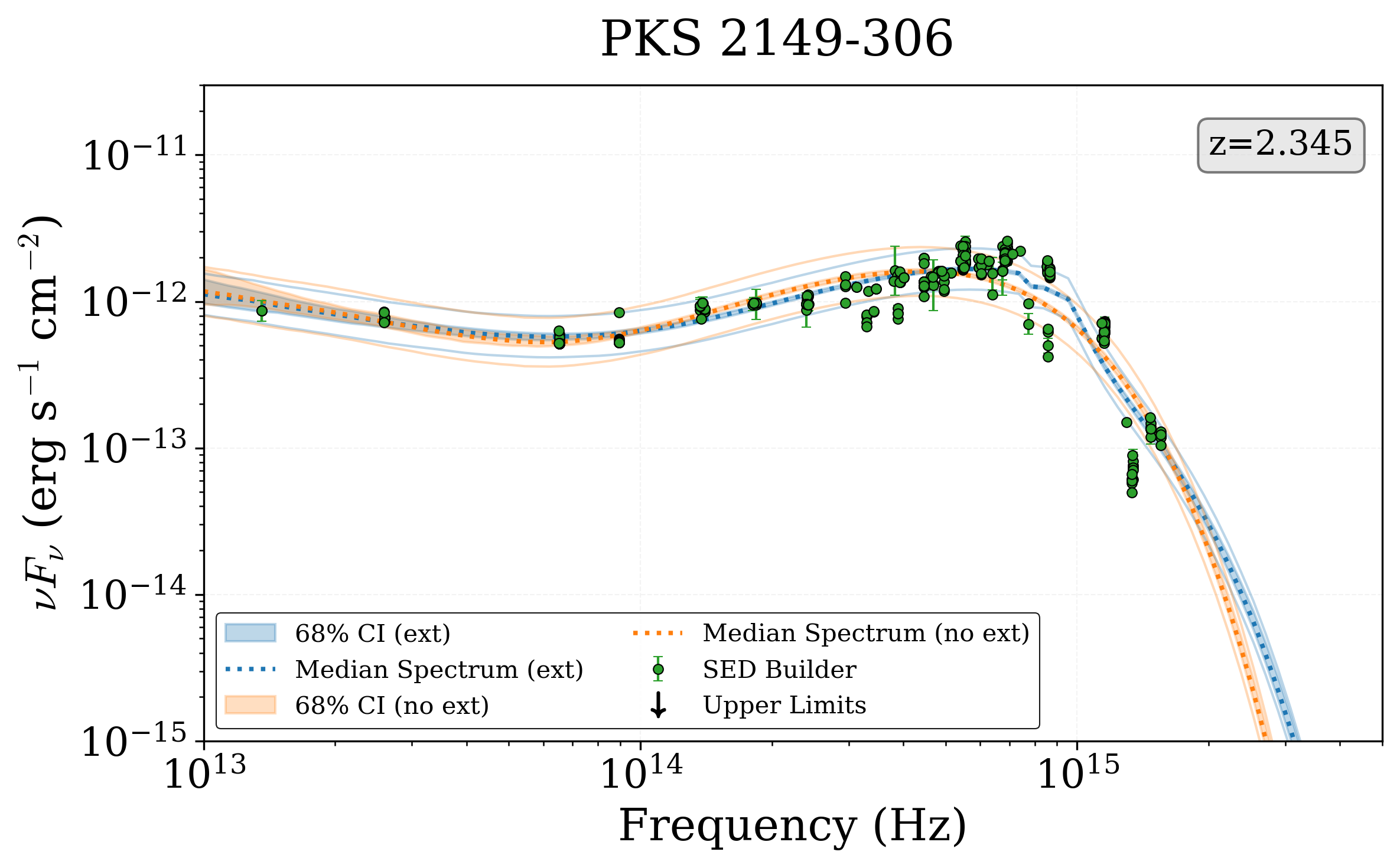}
    \caption{Infrared-to-ultraviolet SEDs of three high-$z$ blazars (in the observer's frame). Green points indicate publicly available photometric observations from the ASI/SSDC SED Builder \citep{Stratta2011}. Upper limits are plotted with black arrows. Model fits with and without IGM extinction are plotted with blue and orange colors, respectively. Dark shaded regions indicate the 68 per cent confidence intervals. Thin solid lines indicate the range of model expectations accounting for the intrinsic scatter described by the parameter $\ln f$.
    All sources, except for J064632+445116, are fitted with a two-component (jet+disk) model.}
    \label{fig:sed}
\end{figure*}

\begin{figure*}
    \centering
    \includegraphics[width=0.49\textwidth]{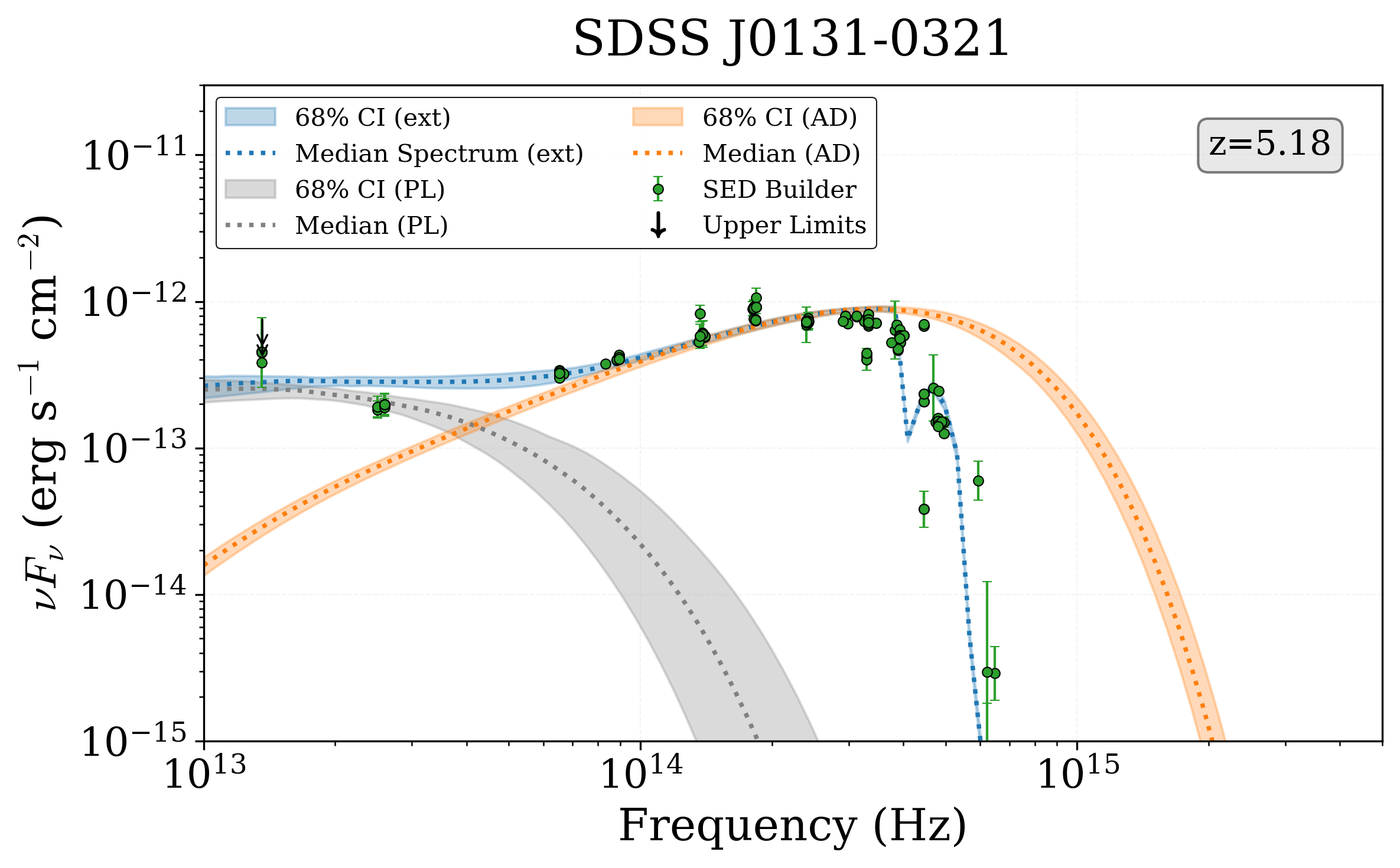}
    \includegraphics[width=0.49\textwidth]{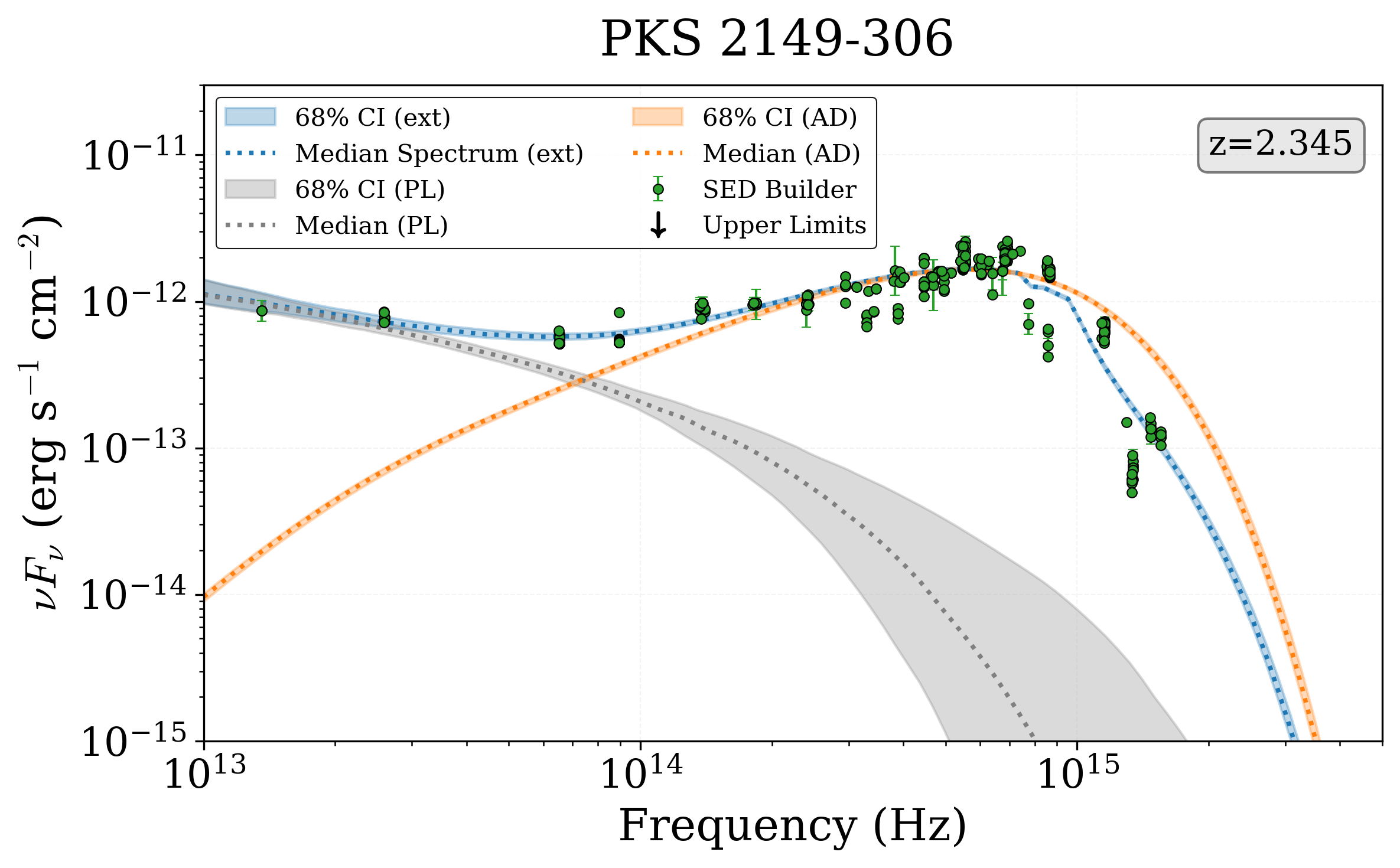}
    \caption{Decomposition of the total (jet+disk) model SEDs presented in Fig~\ref{fig:sed}. For clarity, individual spectral components are plotted without IGM attenuation (power law in gray and multi-temperature blackbody in orange).}
    \label{fig:sed-1}
\end{figure*}
 
\begin{figure*}
    \centering
    \includegraphics[width=0.33\textwidth]{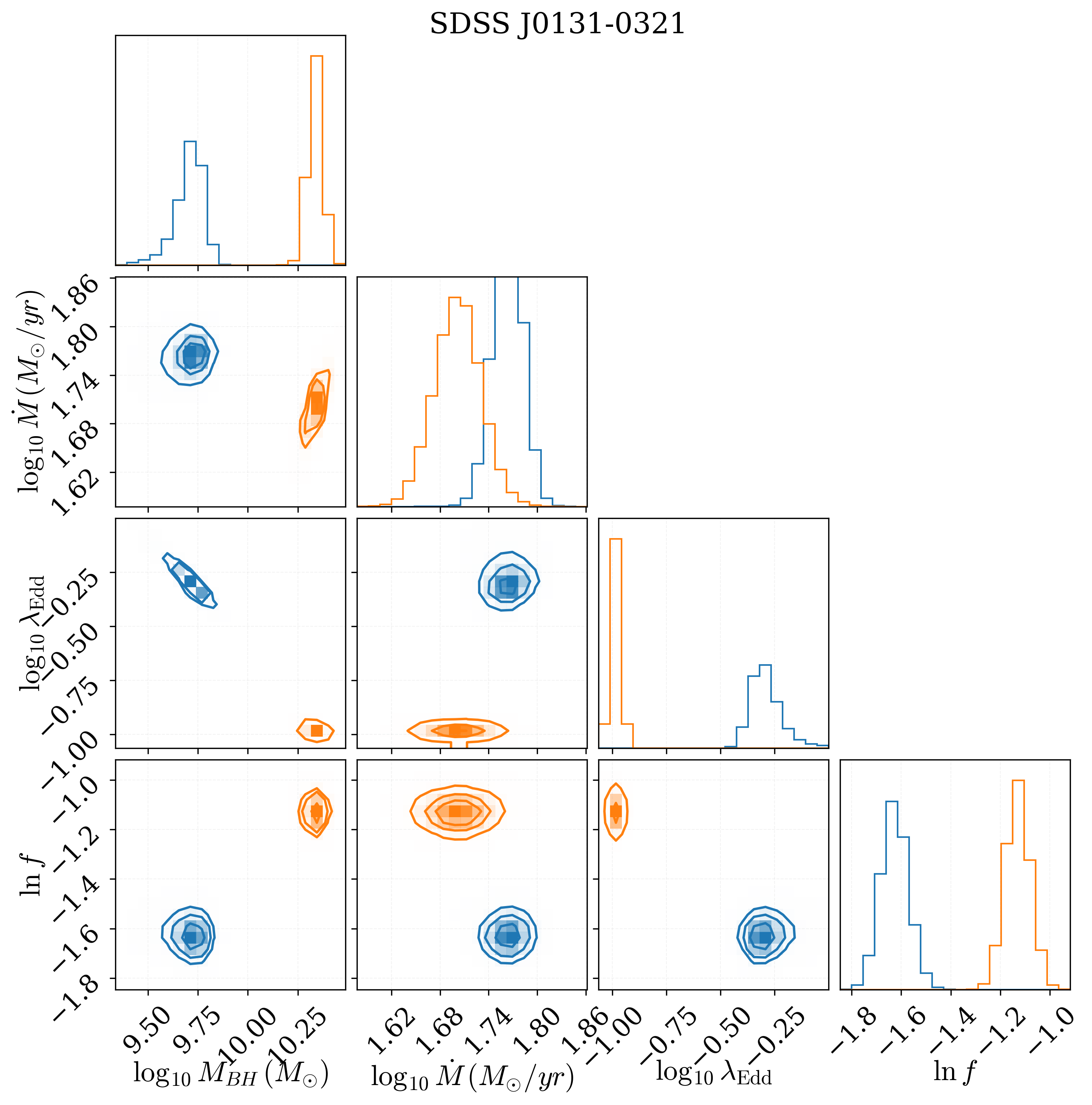}
    \includegraphics[width=0.33\textwidth]{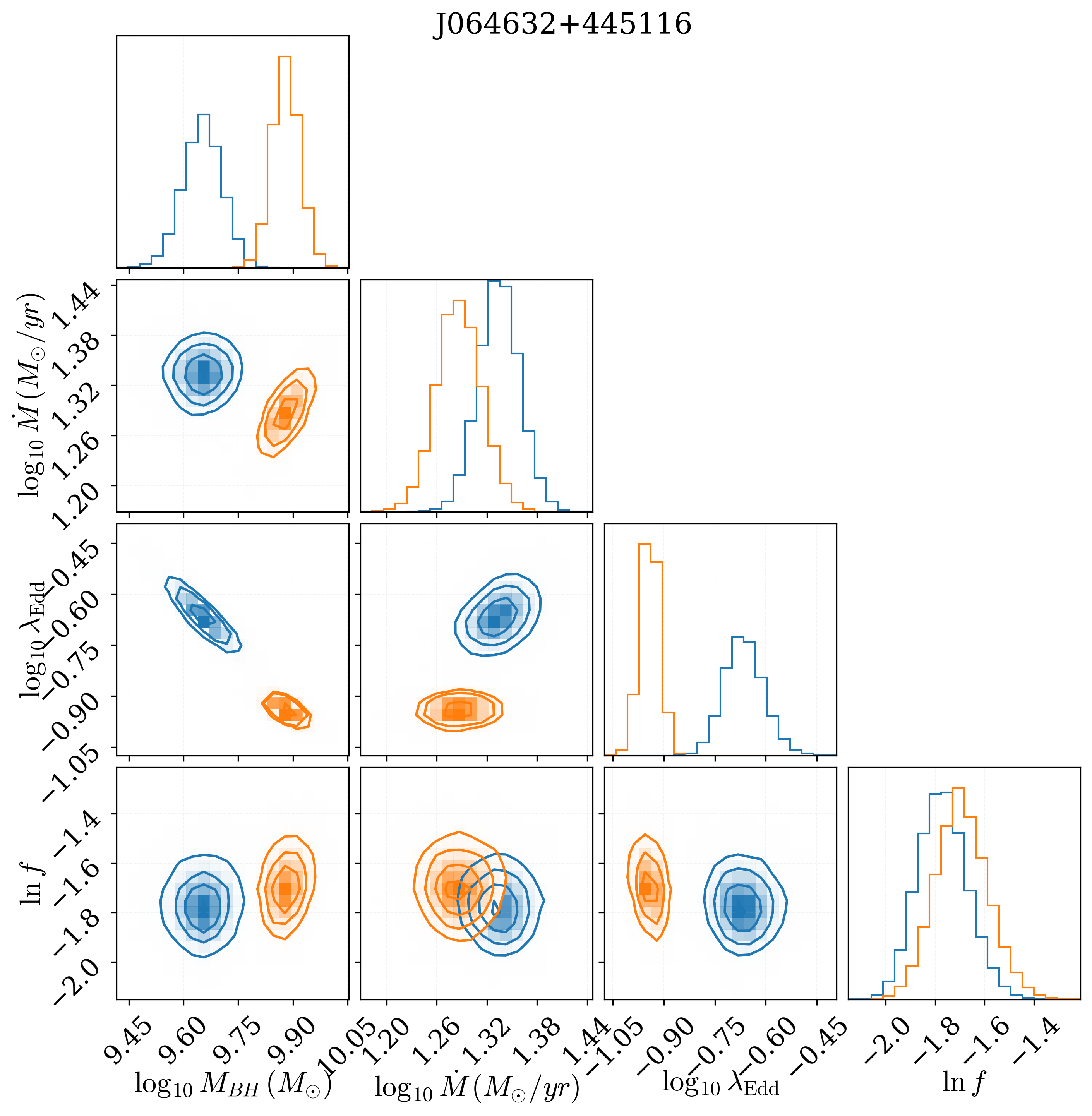}
    \includegraphics[width=0.33\textwidth]{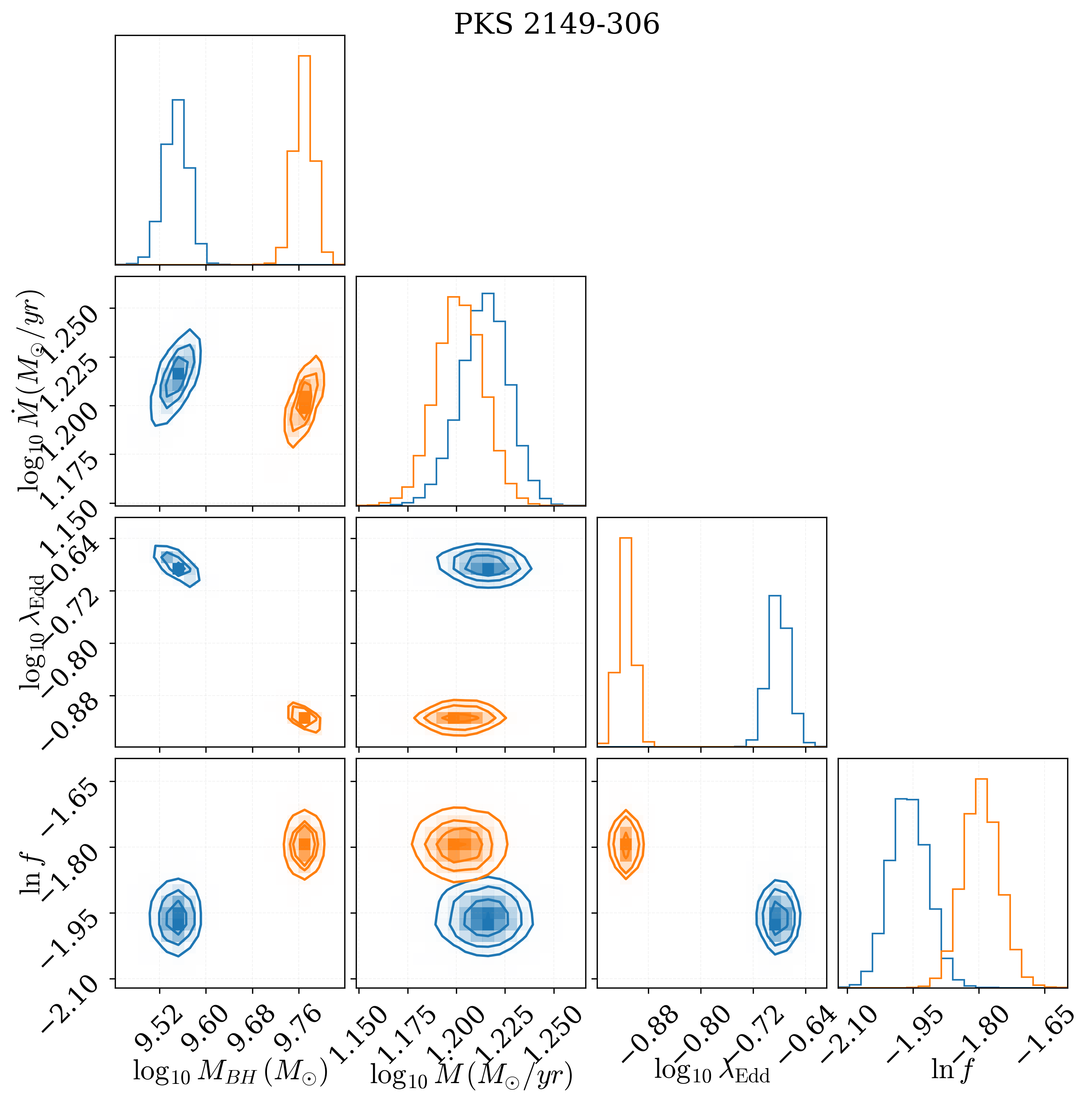}
    \caption{Reduced corner plots for four blazars showing the posterior distributions for the pair of fitted parameters ($M_{\rm BH}, \dot{M}$) and $\ln f$. The distribution of the Eddington ratio, which is a derived quantity, is also shown for comparison. Colors have the same meaning as in Fig.~\ref{fig:sed}.}
    \label{fig:corner}
\end{figure*}

\section{Results} \label{sec:results}
The MCMC fitting results are presented in Tables~\ref{tab:fit_params} and \ref{tab:fit_params_power_law}. More than half of the sample (16 of the 23 sources) was fitted with a two-component (jet+disk) model. The black hole masses lie in the range $3 \times 10^8 M_\odot - 10^{10}~M_\odot$, and the inferred mass accretion rates are $2 \, M_\odot/{\rm yr} \lesssim \dot{M} \lesssim 200 \, M_\odot/\rm yr$. We find no strong correlation between the black hole mass and mass accretion rates. However, we find an empirical relation between the two variables, taking into account the likely scatter introduced by the unknown black hole spin (for more details, see Appendix~\ref{appD}).

We present indicative results of our MCMC fit to the photometric SEDs of three high-$z$ blazars from the sample\footnote{SEDs and corner plots for all sources are available on this \href{https://zenodo.org/records/17781963}{link}.}. The SED and reduced corner plots are shown, respectively, in Figs.~\ref{fig:sed} and \ref{fig:corner}. A decomposition of the jet+disk SED model is also shown in Fig.~\ref{fig:sed-1}. Inspection of the corner plots reveals that fits without accounting for extinction due to hydrogen gas in the IGM lead to an overestimation of the black hole masses (and underestimation of the Eddington ratios), which generally becomes larger for higher redshift sources. The effects of extinction on the parameter estimation across the sample are presented in Appendix~\ref{appC}. 

The three objects were selected to highlight the distinct capabilities of the MCMC SED fitting procedure. SDSS~J0131-0321 is included as a high-redshift case to demonstrate the method performance under strong IGM attenuation and to verify recovery of the disk peak at large cosmological distances. J064632+445116 exemplifies the case where a single spectral component can describe the data. Moreover, its SED includes high-frequency upper limits, thus showcasing the likelihood treatment of censored data and illustrating how upper limits influence the posterior without biasing mass inference. Finally, PKS~2149-306 was selected as a representative well-sampled source at lower redshifts. It has dense multi-wavelength coverage and satisfactorily converged MCMC chains (especially when IGM extinction is accounted for), providing a prototypical comparison for both complex and data-limited fits. 

\begin{figure*}[htbp]
    \centering
    \includegraphics[width=0.45\textwidth]{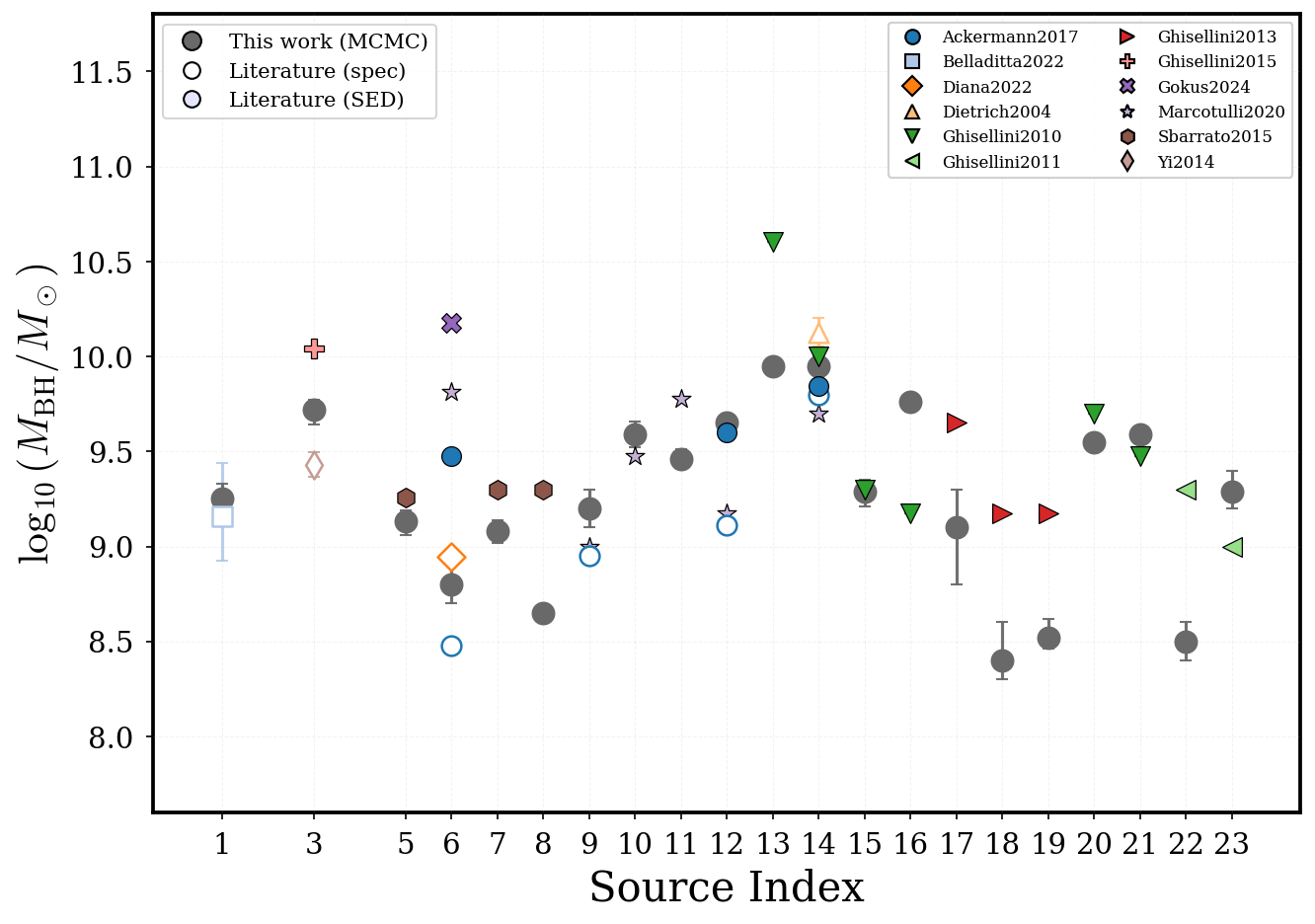}
    \includegraphics[width=0.45\textwidth]{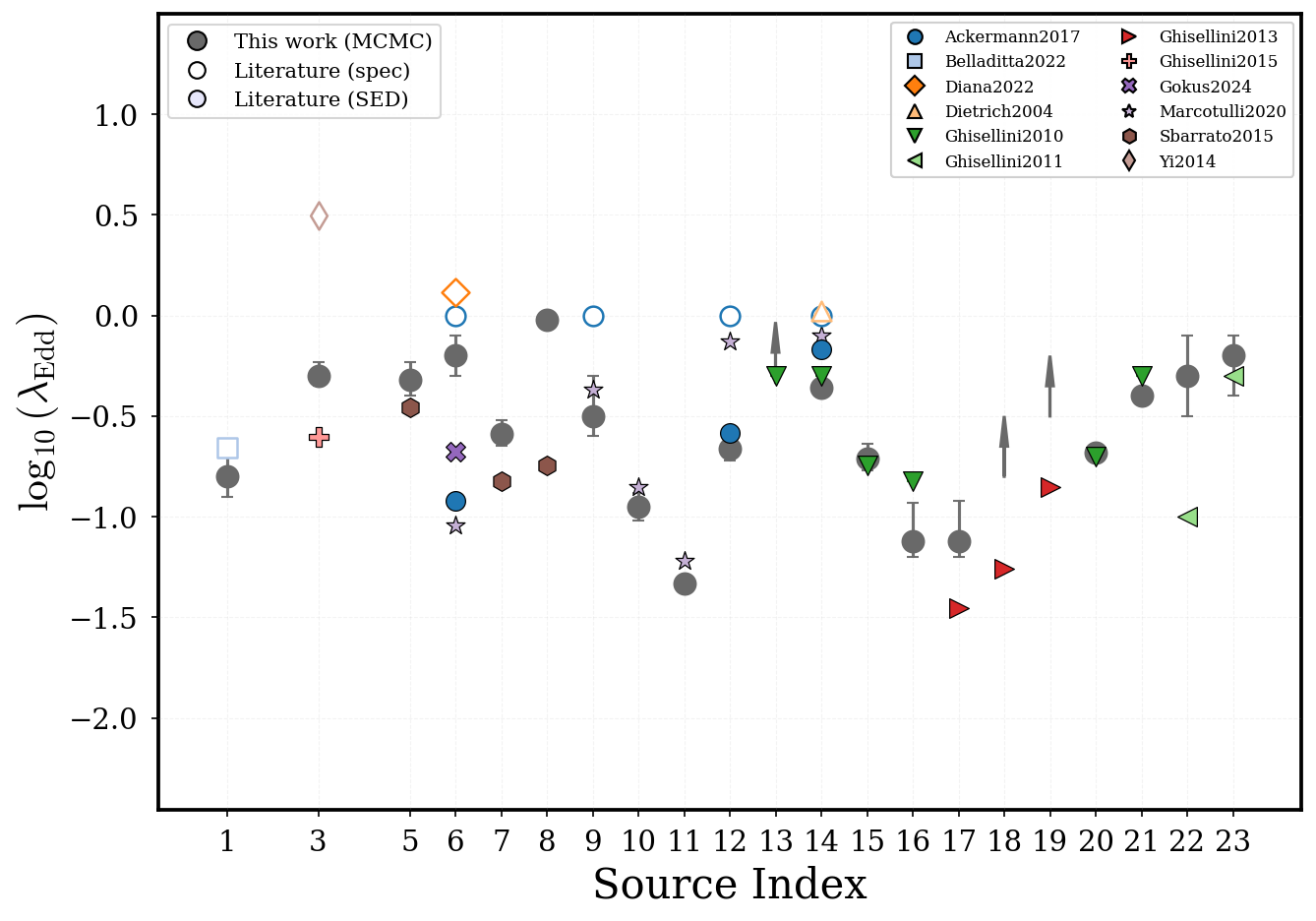}
    \makebox[\textwidth][r]{%
        \begin{minipage}[t]{0.47\textwidth}
            \includegraphics[width=\linewidth]{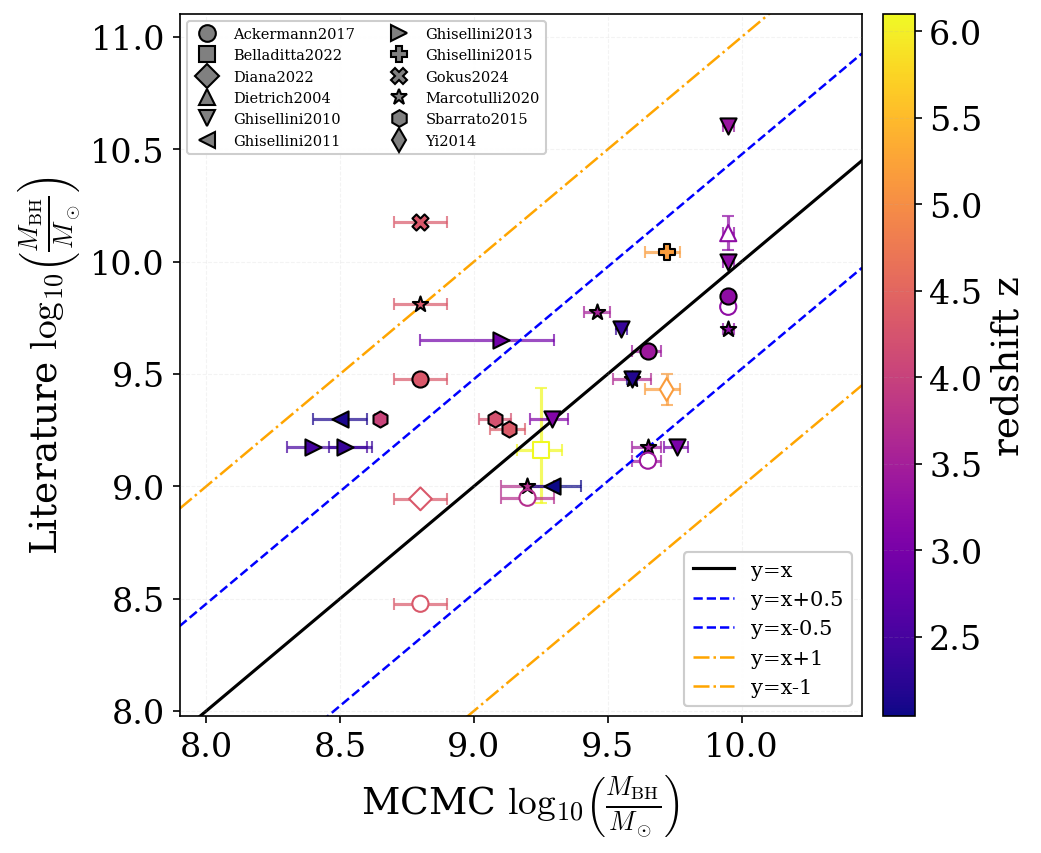}
        \end{minipage}%
        \hspace{0.01\textwidth}%
        \begin{minipage}[t]{0.47\textwidth}
            \includegraphics[width=\linewidth]{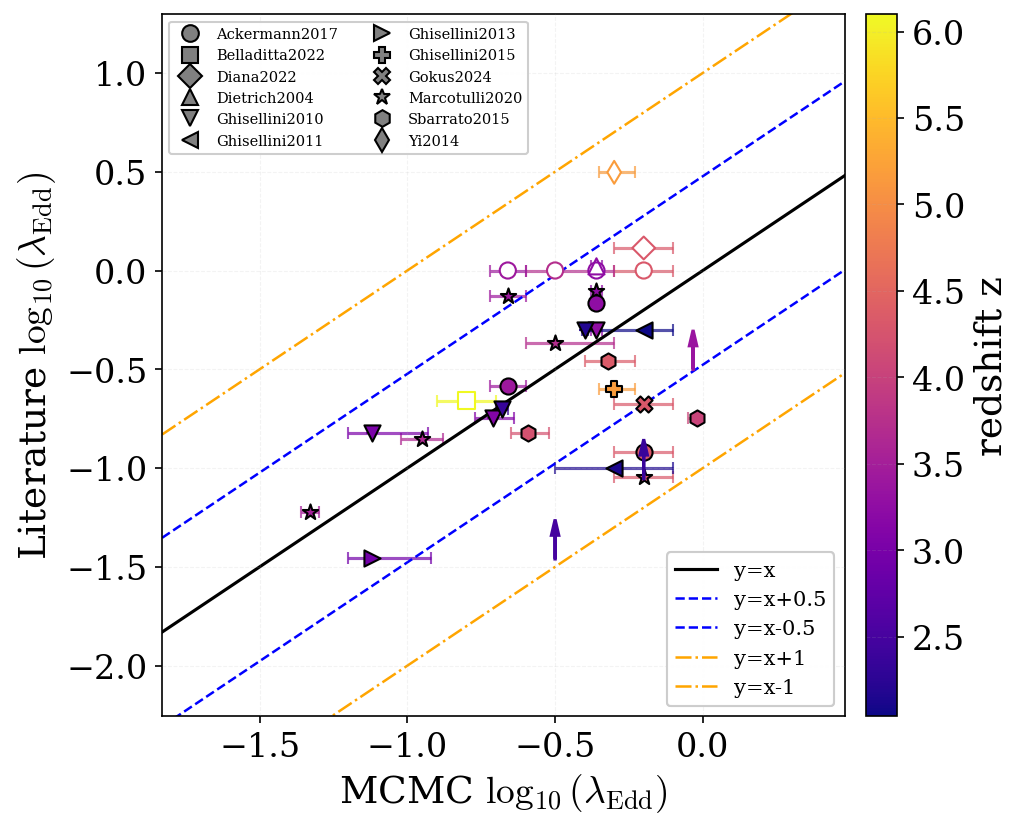}
        \end{minipage}%
        }
\caption{Comparison between our MCMC estimates (including extinction) and literature values for black hole mass $M_{\rm BH}$ (left column) and Eddington    ratio $\lambda_{\rm Edd}$ (right column). 
    \textit{Top left panel:} Black hole mass estimates for each source (index on x-axis) from our MCMC (black circles) and literature (colored markers, with open symbols for spectroscopic and filled for SED-fitting estimates). Error bars show 68\% uncertainties. Marker shapes indicate different literature sources.
    \textit{Top right panel:} Eddington ratio comparison, maintaining identical visual coding to left panel.
    \textit{Bottom left panel:} Direct comparison of black hole mass estimates between our MCMC method (x-axis) and literature values (y-axis), with points colored by source redshift $z$ and marker shapes indicating literature sources. Open symbols represent spectroscopic measurements while filled symbols denote SED-fitting estimates. MCMC uncertainties (68\% confidence) are displayed.
    \textit{Bottom right panel:} Identical in format to bottom left panel, presenting Eddington ratio comparisons with the same redshift coloring and marker coding system.
    In the bottom panels we overplot lines to guide the eye (see inset legend). Blazars B2~1023+25 and GB6 B1428+4217 are not shown for reasons explained in Appendix \ref{sec:Results_Fit}. Upward-pointing arrows indicate lower limits on the MCMC Eddington ratio estimates for the sources marked in \ref{tab:fit_params}.
}
    \label{fig:mass_edd_comparison}
\end{figure*}

Figure~\ref{fig:mass_edd_comparison} presents a compact four-panel comparison between our MCMC-derived estimates (including extinction) and values from the literature for the black-hole mass $M_{\rm BH}$ (left column) and the Eddington ratio $\lambda_{\rm Edd}$ (right column). Error bars indicate the 68\% range of values from the posterior distributions. The uncertainties in the Eddington ratio are typically larger than those in the black hole mass, as they also incorporate the uncertainty in the estimation of the mass accretion rate. All values taken from the literature, except three measurements obtained with spectroscopic methods (see also Table~\ref{tab:sample}), have no reported uncertainties.  

The top row presents source-by-source comparisons, with each object positioned at its original index on the x-axis. Points are grouped by literature source (each reference is represented by a distinct marker shape and color) to test for paper-specific systematics. Our MCMC estimates are shown as black circles, while literature values use colored markers with open symbols for spectroscopic measurements and filled symbols for SED-fitting results. MCMC asymmetric 68\% confidence intervals are displayed. Visual inspection of the top panels reveals no evidence of systematic offsets associated with particular studies, as points from each reference are distributed around our MCMC values without consistent clustering. Approximately half of the sample exhibits black hole mass estimates from literature that are higher than our MCMC results, with discrepancies reaching up to a factor of five in some cases. Notably, for sources where both spectroscopic and SED-based mass estimates are available in the literature, our MCMC results typically align more closely with the spectroscopic measurements. The bottom row uses identical marker shapes for a given literature group, but maps the marker color to the redshift. Trends with marker color could reveal biases that grow with $z$ (for example, stronger IGM attenuation, reduced photometric coverage of the disk peak, or calibration differences between optical and IR surveys that dominate at different redshifts). We find no indication of redshift-dependent trends in either black hole mass or the Eddington ratio. Interestingly, we find blazars accreting at high Eddington ratios, $\lambda_{\rm Edd} > 0.4$, both at $z \sim 2-3$ and $z\sim 5-6$.

\section{Discussion}\label{sec:discussion}
In our analysis, we hypothesized that the accretion disk extends down to six gravitational radii away from the black hole, which is equivalent to assuming non-spinning black holes in all sources. In Sect.~\ref{sec:spin} we discuss how a non-zero spin of the black hole would affect our results. Moreover, in Sect.~\ref{sec:seeds} we discuss the implications of our results on the black hole seeds of high$-z$ blazars in the framework of the Eddington-limited black hole growth. 

\subsection{Effect of black-hole spin on the thin-disk spectrum}\label{sec:spin} 
For the purposes of this discussion we adopt the standard multi-temperature blackbody prescription with a zero-torque inner boundary and the inner disk edge located at the innermost stable circular orbit (ISCO), which depends on the dimensionless Kerr spin parameter \(a\) (\citealt{1973A&A....24..337S,Bardeen1972,1973blho.conf..343N}). For a Kerr black hole, the ISCO radius is given by
\begin{align}
\label{eq:Kerr}
\frac{R_{\rm ISCO}}{R_{\rm g}} &= 3 + Z_2(a) - \mathrm{sign}(a)\sqrt{(3 - Z_1(a))(3 + Z_1(a) + 2Z_2(a))}, \\
Z_1(a) &= 1 + (1 - a^2)^{1/3}\Big[(1+a)^{1/3} + (1-a)^{1/3}\Big], \\
Z_2(a) &= \sqrt{3a^2 + Z_1(a)^2}.
\end{align}
The local effective temperature profile is also given by Eq.~\ref{eq:Tr}
where \(R_{\rm in} = R_{\rm ISCO}\) (the relativistic correction factors, which are discussed in \citealt{1973blho.conf..343N},  are not applied here). 

The maximum disk temperature, which is attained near \(r \approx R_{\rm in}\), scales as $T_{\max} \propto (\dot{M} M_{\rm BH})^{1/4} R_{\rm in}^{-3/4}$. The peak frequency, $\nu_{\rm peak}$, of the accretion disk spectrum in units of \(\nu F_\nu\) is \(\nu_{\rm peak} \propto T_{\rm max}\), and the corresponding peak flux scales as $(\nu F_\nu^{\rm AD})_{\max} \propto T_{\max}^4 R_{\rm in}^2$. From these scalings, the effect of changing spin can be exactly compensated for if both $M_{\rm BH}$ and $\dot M$ are adjusted simultaneously. 

Let primed and unprimed quantities indicate the variables of a black hole with spin $a\neq 0$ and $a=0$, respectively. Imposing the same (observed) peak frequency and peak flux, we derive the following relations:
\begin{align}
\left(\frac{\dot M'}{M_{\rm BH}'^{\,2}}\right)^{1/4} R_{\rm in}'^{-3/4} 
&=\left(\frac{\dot M}{M_{\rm BH}^{2}}\right)^{1/4} R_{\rm in}^{-3/4}, \\
\dot M'\,R_{\rm in}'^{-1}&=\dot M\,R_{\rm in}^{-1}.
\end{align}
which when combined yield the rescaling of the black hole mass and accretion rate:
\begin{align}
\label{eq:scale}
M_{\rm BH}'=M_{\rm BH}\,\frac{R_{\rm in}}{R_{\rm in}'},
\qquad
\dot M'=\dot M\,\frac{R_{\rm in}'}{R_{\rm in}}.
\end{align}
For example, changing from $a=0$ ($R_{\rm in}=6$) to $a=0.998$ ($R_{\rm in}\simeq1.24$) would result in
\begin{equation}
M_{\rm BH}'\simeq 4.8\,M_{\rm BH},\qquad
\dot M'\simeq 0.2\,\dot M,
\label{eq:rescale}
\end{equation}
in order to keep \(\nu_{\rm peak}\) and $(\nu F_\nu^{AD})_{\max}$ unchanged. 

\begin{figure}
\centering
\includegraphics[width=0.45\textwidth]{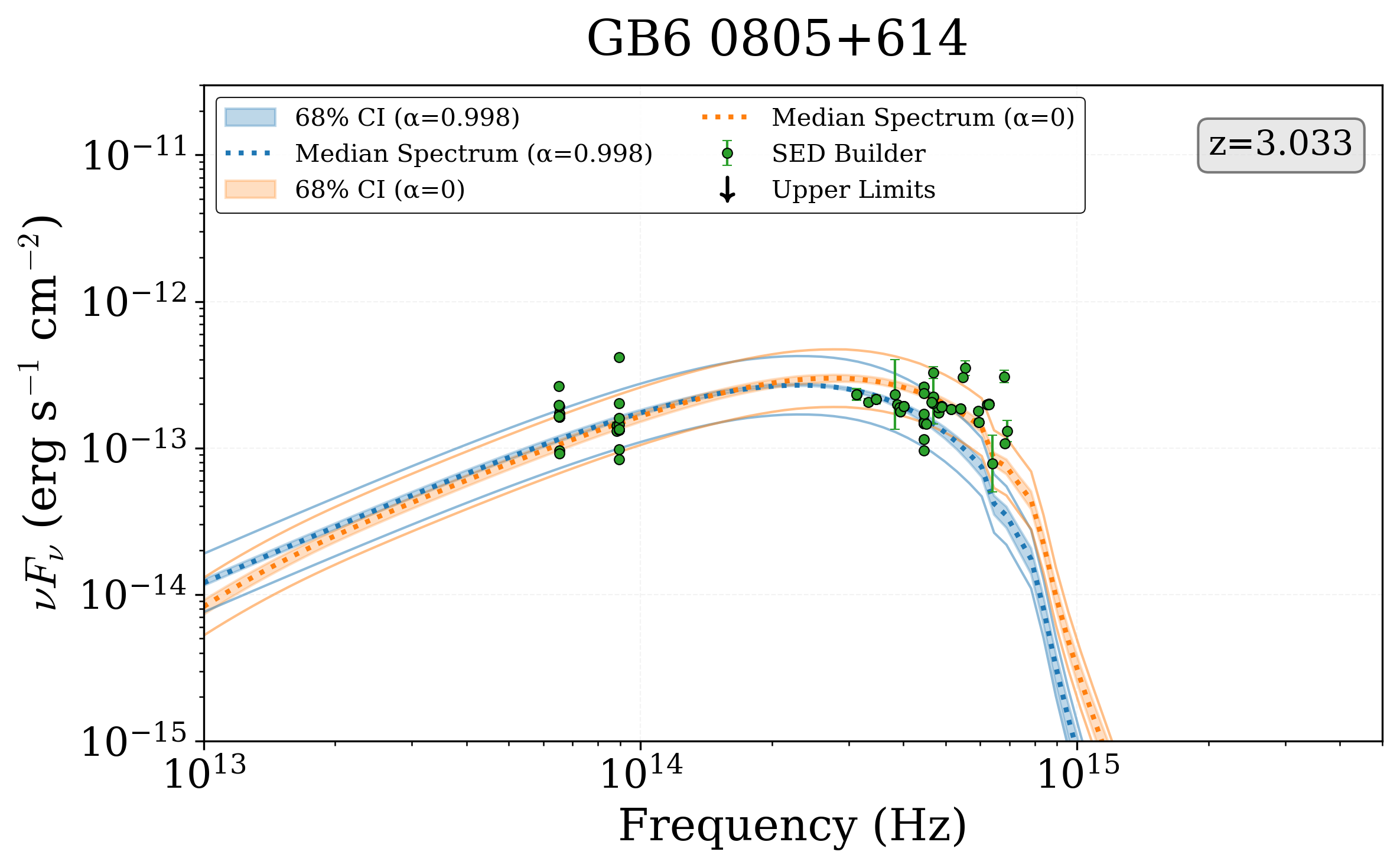}
\includegraphics[width=0.45\textwidth]{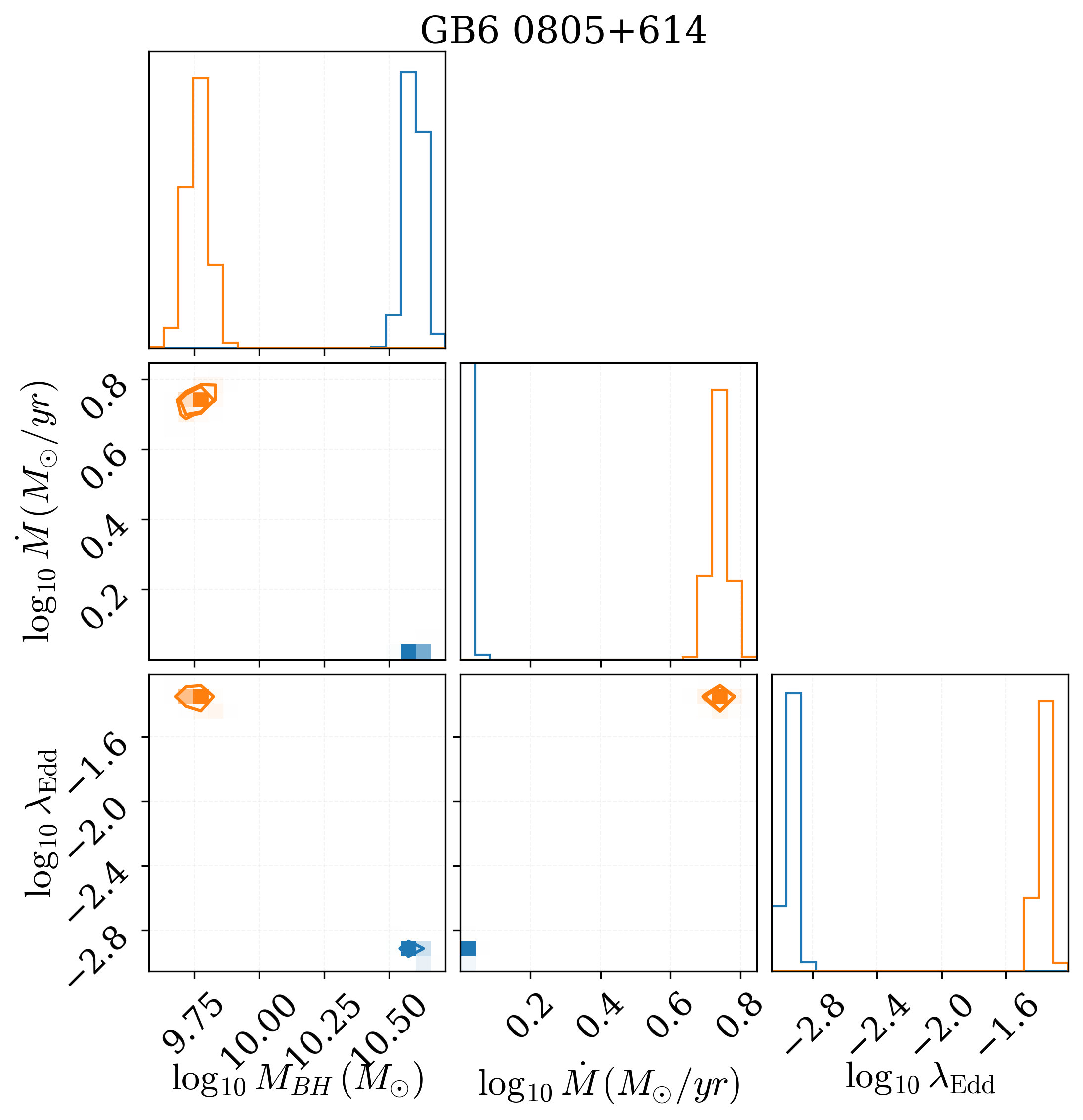}
\caption{Effects of black hole spin on SED disk fitting results. \textit{Top panel:} SED of GB6~0805+614 with MCMC modeling results for a black hole with $a=0.998$ and $a=0$ overplotted in blue and orange colors, respectively. \textit{Bottom panel:} Corner plot showing the posterior distributions for $\dot{M}, M_{\rm BH}$ and $\lambda_{\rm Edd}$ for a black hole with $a=0.998$ and $a=0$. Same color coding as in the top panel is used. Different choices of $(M_{\rm BH}, \dot{M})$, depending on the black hole spin, lead to the same accretion disk spectrum.}
\label{fig:spin}
\end{figure}

To illustrate the impact of black hole spin on our results, we chose one source from the sample GB6~0805+614,  where extinction is not very important, and repeated the MCMC fit assuming a maximally spinning black hole. The results are presented in Fig.~\ref{fig:spin}. The SEDs are identical to those of a non-spinning black hole, if the black hole is more massive and accretes at a lower rate, in agreement with the scalings of Eq.~\ref{eq:rescale}. The Eddington ratio for the solution of a maximally spinning black hole is $\approx 0.04$ times smaller than the Eddington ratio for the non-spinning black-hole solution. In this particular example, $\dot{M}'/\dot{M}'_{\rm Edd} < 0.01$ for $a = 0.998$, where the validity of the adopted radiatively efficient disk model breaks down \citep[e.g.][]{1995ApJ...452..710N}. 

Our current disk model does not include general relativistic effects, which become increasingly important for rapidly spinning black holes. Recently, \cite{2023MNRAS.525.3455H}  introduced {\tt RELAGN}, a public code that accounts for both special and general relativistic effects. To explore the impact of relativistic effects, we applied {\tt RELAGN} to GB6~0805+614. We searched for parameters that produce the same peak flux and position as those of our best-fit SEDs shown in Fig.~\ref{fig:spin}. For $\alpha=0$ and 0.998 we found $M_{\rm BH}=4\times 10^{9}~M_\odot$ and $10^{10}~M_\odot$,  with corresponding Eddington ratios of $\log_{10}\lambda_{\rm Edd}=-1.84$ and $-1.9$, respectively. Therefore, accounting for relativistic effects in disk modeling reduces the discrepancies in the inferred black-hole mass and nearly eliminates the dependence of the Eddington ratio on black-hole spin.

To summarize, black hole spin introduces a systematic degeneracy: a disk formed around a high–spin black hole can reproduce the same SED peak frequency and flux as a disk around a low–spin black hole by trading off $M_{\rm BH}$ and $\dot M$ according to the scalings above. In practice, this means that estimates of $M_{\rm BH}$ and $\dot M$ from SED disk fitting can be biased by factor of up to five (the Eddington ratios up to a factor of 25) depending on the assumed spin, unless external constraints are used. For example, the luminosity of the BLR can provide a prior on the disk luminosity through the relation $L_{\rm BLR} \simeq (f_{\rm BLR}/0.1)\,L_{\rm disk}$, but this approach introduces additional uncertainties associated with the covering fraction and the conversion from line to bolometric luminosity. However, we note that the differences found between the literature values and the MCMC fit results of this work (shown in Fig.~\ref{fig:mass_edd_comparison}) cannot be attributed to the black hole spin, as these studies also worked under the hypothesis of a zero-spin black hole.

\subsection{Black-hole seed masses}\label{sec:seeds}
\label{subsec:seed_inversion}
We use the Eddington-limited black hole growth solution for radiatively efficient accretion to derive strict lower bounds on the seed masses required to assemble the black holes observed in our sample. 

Let \(M(t)\) denote the black hole mass at cosmic time \(t\)\footnote{Cosmic time \(t\) and redshift \(z\) are related by \citep{Hogg1999,Planck2016}:
\[
t(z)=\frac{1}{H_0}\int_z^{\infty}\frac{dz'}{(1+z')\sqrt{\Omega_{\rm m}(1+z')^3+\Omega_{\Lambda}}}\;.
\]}, \(\eta\) the radiative efficiency of the disk, and \(\lambda\equiv\lambda_{\rm Edd}\) the Eddington ratio. The mass growth rate is
\begin{equation}
  \frac{dM}{dt}=\frac{(1-\eta)\,L_{\rm acc}}{\eta c^2}
  =\frac{(1-\eta)}{\eta}\,\frac{\lambda L_{\rm Edd}}{c^2},
\end{equation}
and, using \(L_{\rm Edd}=(4\pi G m_p c/\sigma_{\rm T})M\), we obtain
\begin{equation}
  \frac{dM}{dt}=M\,\frac{\lambda(1-\eta)}{\tau_S},
  \qquad
  \tau_S \equiv \frac{\eta\,\sigma_{\rm T}\,c}{4\pi G m_p} \simeq 45~\eta_{-1}~\rm Myr.
  \label{eq:tauS_def_final}
\end{equation}
Here \(\tau_S\) is the Salpeter time \citep{Salpeter1964}, the characteristic e-folding scale for Eddington-limited growth. Assuming constant $\eta$ and $\lambda$, integration yields
\begin{equation}
  M(t)=M_{\rm seed}\exp\!\left[\frac{\lambda(1-\eta)}{\tau_S}(t-t_0)\right],
\end{equation}
which can be inverted to give the seed mass at formation time \(t_0\) required to reach \(M_{\rm final}\equiv M(t_{\rm obs})\) by observation time \(t_{\rm obs}\),
\begin{equation}
  M_{\rm seed}=M_{\rm final}\exp\!\left[-\frac{\lambda(1-\eta)}{\tau_S}(t_{\rm obs}-t_0)\right].
  \label{eq:Mseed_final}
\end{equation}

Equation~\eqref{eq:Mseed_final} thus returns the mathematically minimal seed mass consistent with constant radiative efficiency, uninterrupted accretion at fixed \(\lambda\), no black-hole mergers, and no extended super-Eddington phases. If accretion is intermittent, \(\lambda\) should be replaced by \(\lambda_{\rm eff}=f_{\rm duty}\lambda\), where \(f_{\rm duty}\) is the accretion duty cycle (the fraction of cosmic time during which the black hole accretes at the assumed Eddington ratio). Values of \(f_{\rm duty}<1\), or the inclusion of mergers, episodic super-Eddington phases, or evolving spin-dependent efficiencies, would relax the strict lower bounds derived here \citep{Volonteri2010,Valiante2016,Inayoshi2020}.

\begin{figure*}[htbp]
    \centering
    \includegraphics[width=0.32\textwidth]{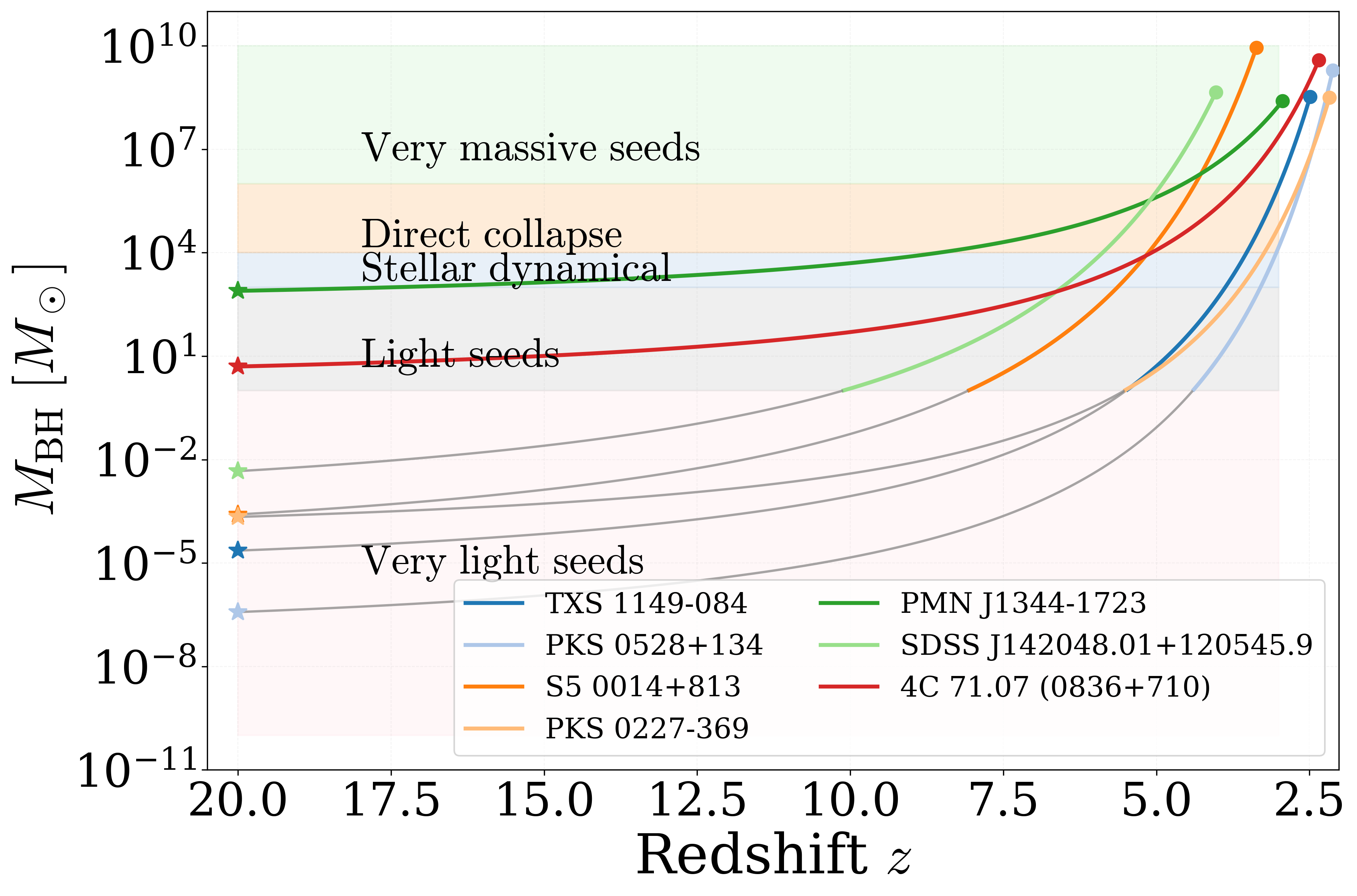}
    \includegraphics[width=0.32\textwidth]{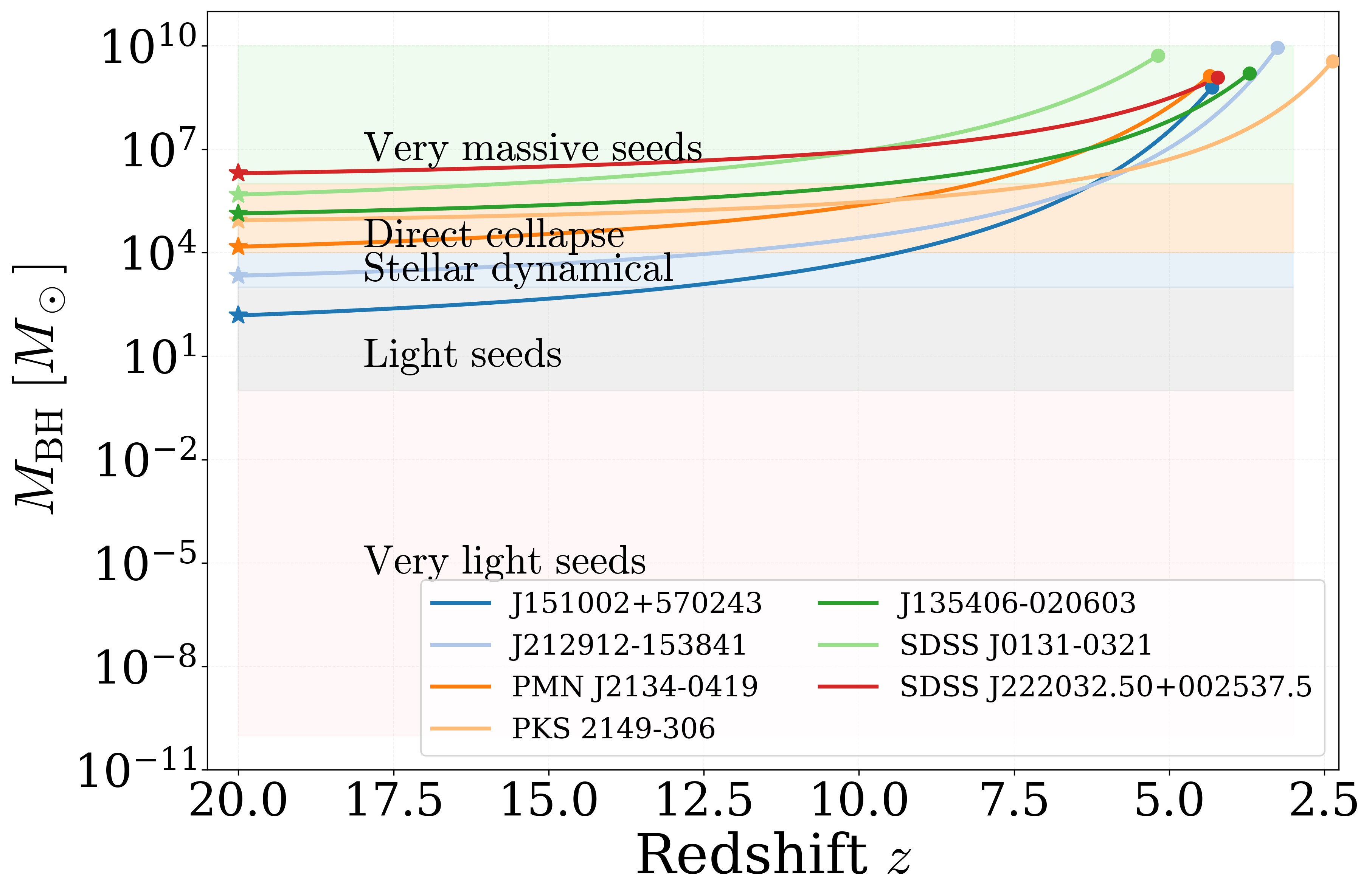}
    \includegraphics[width=0.32\textwidth]{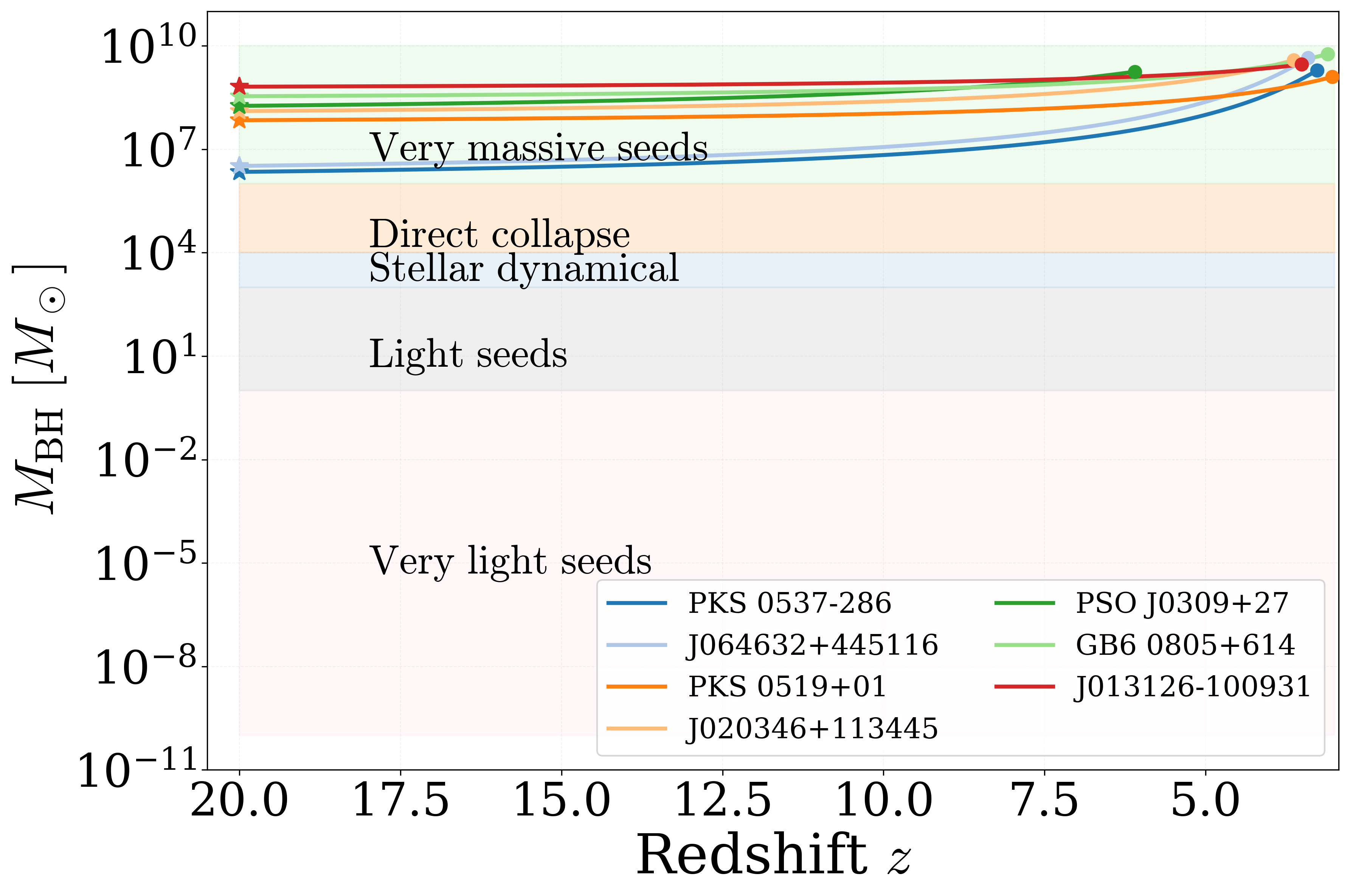}    
    \caption{Black hole growth history for the investigated objects. 
    The shaded horizontal regions correspond to different black hole seeding channels: 
    Pop~III remnants ($M_{\rm seed} \lesssim 10^2\,M_\odot$, grey), stellar dynamical 
    processes ($M_{\rm seed} \sim 10^3$--$10^4\,M_\odot$, cyan), and direct collapse 
    ($M_{\rm seed} \sim 10^4$--$10^6\,M_\odot$, orange). These seed mass ranges are 
    based on the classification described in \citet{Valiante2016}. Parts of tracks with 
    $M_{\rm seed} < 1\,M_\odot$ are shown in light grey, as such low masses are not 
    expected to form black holes.}
    \label{fig:growth_curves}
\end{figure*}

Table~\ref{tab:seed_inversion_ext} presents the inversion results for 21 blazars from our sample. We adopt the median values for \(M_{\rm final}\) and for the Eddington ratio \(\lambda_{\rm Edd}\) from the MCMC fits of this work, assume seed formation at \(z_{\rm seed}=20\) (cosmic time \(t_0\)), and fix \(\eta=0.1\). The table reports the observed redshift \(z_{\rm obs}\), \(\lambda_{\rm Edd}\), \(M_{\rm final}\), the available growth interval \((t_{\rm obs}-t_0)\), the inferred seed mass, and a classification according to seeding channel. We adopt the following literature-motivated bands: \emph{very light seeds}, \(M_{\rm seed}<1\,M_\odot\); Pop\,III remnants (light seeds), \(1\!-\!10^{3}\,M_\odot\)\citep{Volonteri2010,Inayoshi2020}; stellar-dynamical or cluster-runaway seeds, \(10^{3}\!-\!10^{4}\,M_\odot\)\citep{Inayoshi2020}; direct-collapse black holes (DCBHs), \(10^{4}\!-\!10^{6}\,M_\odot\)\citep{Valiante2016,Inayoshi2020}; and \emph{very massive seeds}, \(M_{\rm seed}\geq10^{6}\,M_\odot\). In particular, the label ``very light seed'' is introduced to flag formal minima below one solar mass, which are not physical (they cannot arise from stellar evolution), and therefore demand cautious interpretation. As a visual complement to Table~\ref{tab:seed_inversion_ext}, Fig.~\ref{fig:growth_curves} shows the reconstructed growth tracks  of representative objects compared against the expected seed formation channels.

\begin{table*}[htbp]
\centering
\caption{Minimum seed masses $M_{\rm seed}$ inferred by inverting Eq.~\eqref{eq:Mseed_final} for the 21 objects in the sample (extinction-corrected fits). Inversion assumes seed formation at $z_{\rm seed}=20$ and $\eta=0.1$. The final mass $M_{\rm final}$ and $\lambda_{\rm Edd}$ are the median values of the posterior distributions from the MCMC fits (with extinction included). The last column gives an assignment to common seed channels by mass. The reported values are rounded to one decimal place, except the Eddington ratios that are rounded to two decimal places. Blazars B2~1023+25 and GB6 B1428+4217 are omitted for reasons discussed in Appendix \ref{sec:Results_Fit}.}
\small
\label{tab:seed_inversion_ext}
\begin{tabular}{l c c c c c c}
\hline\hline
Object & $z_{\rm obs}$ & $\lambda_{\rm Edd}$ & $M_{\rm final}\ [M_\odot]$ & $M_{\rm seed}\ [M_\odot]$ & Growth time [Gyr] & Channel \\
\hline
PKS 0528+134 & $2.12$ & $0.63$ & $2.0\times10^{9}$ & $3.8\times10^{-7}$ & $2.9$ & Very light seeds -- unphysical \\
TXS 1149-084 & $2.49$ & $0.63$ & $3.3\times10^{8}$ & $2.3\times10^{-5}$ & $2.4$ & Very light seeds -- unphysical \\
PKS 0227-369 & $2.17$ & $0.50$ & $3.2\times10^{8}$ & $2.2\times10^{-4}$ & $2.8$ & Very light seeds -- unphysical \\
S5 0014+813 & $3.37$ & $0.93$ & $8.9\times10^{9}$ & $2.5\times10^{-4}$ & $1.7$ & Very light seeds -- unphysical\\
SDSS J142048.01+120545.9 & $4.03$ & $0.96$ & $4.5\times10^{8}$ & $4.6\times10^{-3}$ & $1.3$ & Very light seeds -- unphysical \\
4C 71.07 (0836+710) & $2.35$ & $0.40$ & $3.9\times10^{9}$ & $5.0\times10^{0}$ & $2.6$ & Light seeds ($1\!-\!10^{3}\,M_\odot$) \\
J151002+570243 & $4.31$ & $0.63$ & $6.3\times10^{8}$ & $1.5\times10^{2}$ & $1.2$ & Light seeds ($1\!-\!10^{3}\,M_\odot$) \\
PMN J1344-1723 & $2.94$ & $0.32$ & $2.5\times10^{8}$ & $7.8\times10^{2}$ & $2.0$ & Light seeds ($1\!-\!10^{3}\,M_\odot$) \\
J212912-153841 & $3.26$ & $0.44$ & $8.9\times10^{9}$ & $2.1\times10^{3}$ & $1.7$ & Stellar-dynamical ($10^{3}\!-\!10^{4}\,M_\odot$) \\
PMN J2134-0419 & $4.35$ & $0.48$ & $1.3\times10^{9}$ & $1.5\times10^{4}$ & $1.2$ & Direct-collapse BHs ($10^{4}\!-\!10^{6}\,M_\odot$) \\
PKS 2149-306 & $2.37$ & $0.21$ & $3.5\times10^{9}$ & $8.7\times10^{4}$ & $2.5$ & Direct-collapse BHs ($10^{4}\!-\!10^{6}\,M_\odot$) \\
J135406-020603 & $3.71$ & $0.32$ & $1.6\times10^{9}$ & $1.4\times10^{5}$ & $1.5$ & Direct-collapse BHs ($10^{4}\!-\!10^{6}\,M_\odot$) \\
SDSS J0131-0321 & $5.18$ & $0.50$ & $5.2\times10^{9}$ & $4.8\times10^{5}$ & $0.9$ & Direct-collapse BHs ($10^{4}\!-\!10^{6}\,M_\odot$) \\
SDSS J222032.50+002537.5 & $4.22$ & $0.26$ & $1.2\times10^{9}$ & $2.0\times10^{6}$ & $1.2$ & Very massive seeds ($\geq 10^{6}\,M_\odot$) \\
PKS 0537-286 & $3.27$ & $0.20$ & $2.0\times10^{9}$ & $2.2\times10^{6}$ & $1.7$ & Very massive seeds ($\geq 10^{6}\,M_\odot$) \\
J064632+445116 & $3.41$ & $0.22$ & $4.5\times10^{9}$ & $3.3\times10^{6}$ & $1.7$ & Very massive seeds ($\geq 10^{6}\,M_\odot$) \\
PKS 0519+01 & $3.03$ & $0.08$ & $1.3\times10^{9}$ & $7.0\times10^{7}$ & $1.9$ & Very massive seeds ($\geq 10^{6}\,M_\odot$) \\
J020346+113445 & $3.63$ & $0.11$ & $3.9\times10^{9}$ & $1.3\times10^{8}$ & $1.5$ & Very massive seeds ($\geq 10^{6}\,M_\odot$) \\
PSO J0309+27 & $6.10$ & $0.16$ & $1.8\times10^{9}$ & $1.8\times10^{8}$ & $0.7$ & Very massive seeds ($\geq 10^{6}\,M_\odot$) \\
GB6 0805+614 & $3.10$ & $0.08$ & $5.8\times10^{9}$ & $3.5\times10^{8}$ & $1.9$ & Very massive seeds ($\geq 10^{6}\,M_\odot$) \\
J013126-100931 & $3.51$ & $0.05$ & $2.9\times10^{9}$ & $6.5\times10^{8}$ & $1.6$ & Very massive seeds ($\geq 10^{6}\,M_\odot$) \\
\hline
\end{tabular}
\end{table*}

The formally tiny values of \(M_{\rm seed}\) reported in Table~\ref{tab:seed_inversion_ext} should be regarded as non-physical minima that signal a degeneracy in the sample, uninterrupted Eddington-limited growth assumption embodied in Eq.~\ref{eq:Mseed_final}. Three interpretations can resolve the unphysically low seed mass: (i) the seeds formed at a later cosmic epoch \(z_0\) (hence the available growth time \(\Delta t = t_{\rm obs}-t_0\) is smaller), or (ii) accretion was intermittent, so that the effective growth rate is reduced by a duty cycle \(f_{\rm duty}<1\), or (iii) the black hole was fast spinning, thus the Eddington rate would be lower than the values obtained in this work 
(as discussed in~\ref{sec:spin}). 

To quantify the second possibility, we replaced the fitted Eddington ratio \(\lambda\) by an effective value \(\lambda_{\rm eff}=f_{\rm duty}\,\lambda\). Rearranging Eq.~\ref{eq:Mseed_final} for \(f_{\rm duty}\) and imposing \(M_{\rm seed}=10\,M_\odot\) gives
\[
f_{\rm duty} \;=\; -\frac{\tau_{\rm S}}{\lambda(1-\eta)\,\Delta t}\,
\ln\!\left(\frac{10\,M_\odot}{M_{\rm final}}\right),
\]
where we adopted \(\eta=0.1\) and \(\Delta t\) is the growth interval used in Table~\ref{tab:seed_inversion_ext} (computed assuming seeds form at \(z_0=20\)).

\begin{table*}
\centering
\caption{Observed redshift $z_{\rm obs}$, fitted Eddington ratio $\lambda$, fitted final mass $M_{\rm final}$, duty-cycle $f_{\rm duty}$ required for a $10\,M_\odot$ seed formed at $z_0=20$, and formation redshift $z_0$ required for a $10\,M_\odot$ seed assuming continuous accretion ($f_{\rm duty}=1$). All computations assume $\eta=0.10$ and $\tau_S=4.5\times10^{7}$~yr and use the $M_{\rm final}$ and $\lambda$ values from Table~\ref{tab:seed_inversion_ext}. The reported values are rounded to one decimal place, except the Eddington ratios that are rounded to two decimal places.}
\label{tab:fduty}
\begin{tabular}{l c c c c c}
\hline
Object & $z_{\rm obs}$ & $\lambda_{\rm Edd}$ & $M_{\rm final}$ [$M_\odot$] & $f_{\rm duty}$ (for $z_0=20$, $M_{\rm seed}=10\,M_\odot$) & $z_0$ (for $f_{\rm duty}=1$, $M_{\rm seed}=10\,M_\odot$)\\
\hline
PKS\,0528+134 & $2.12$ & $0.63$ & $2.0\times10^{9}$ & $0.5$ & $4.0$ \\TXS\,1149-084 & $2.49$ & $0.63$ & $3.3\times10^{8}$ & $0.6$ & $4.8$ \\
PKS\,0227-369 & $2.17$ & $0.50$ & $3.2\times10^{8}$ & $0.6$ & $4.7$ \\
S5\,0014+813 & $3.37$ & $0.93$ & $8.9\times10^{9}$ & $0.7$ & $7.0$ \\
SDSS\,J142048.01+120545.9 & $4.03$ & $0.96$ & $4.5\times10^{8}$ & $0.7$ & $8.5$ \\
\hline
\end{tabular}
\end{table*}

The numerical values in Table~\ref{tab:fduty} show that, under the conservative assumptions stated above, these seven objects would require duty cycles 0.5-0.7 if their seeds were \(10\,M_\odot\) and formed at \(z_0=20\). Alternatively, if accretion proceeded continuously (\(f_{\rm duty}=1\)) the required formation redshifts for a \(10\,M_\odot\) seed span a broad range: some sources would be consistent with seeds forming at relatively modest redshifts (\(z_0\sim 4.0\) for PKS\,0528$+$134), while others, such as SDSS J142048.01+120545.9 , would require seeds forming still at high redshift ($z_0 > 6$).

Several important caveats apply. First, the radiative efficiency \(\eta\) enters linearly in the Salpeter time and therefore a larger \(\eta\) increases the required duty cycle or the required formation redshift; hence, uncertainties in black-hole spin and accretion geometry propagate directly into the inferences \citep{Volonteri2010}. Second, mergers and short episodes of super-Eddington accretion (both plausible in galaxy formation) reduce the need for long, continuous Eddington-limited accretion and would relax the constraints in Table~\ref{tab:fduty} significantly \citep{Begelman2006,Volonteri2010}. An additional point of support comes from the semi-analytic and merger-driven calculations of \cite{Yoo2004}, who demonstrated that when hierarchical mergers are included alongside sustained Eddington-limited gas accretion, stellar-remnant seeds of order \(M_{\rm seed}\sim 10\,M_\odot\) can plausibly grow to \(\sim10^{9}\,M_\odot\) by \(z\simeq6.4\); this provides a concrete example in which the ``very light seed'' entries in Table~\ref{tab:fduty} can be reconciled with physically plausible seed masses through the inclusion of mergers and prolonged accretion. Therefore, the label ``very light seed - unphysical'' in Table~\ref{tab:seed_inversion_ext} should be interpreted as a flag that the inversion is exposing a degeneracy between formation epoch and duty cycle rather than implying literal sub-stellar black holes.

For the objects labeled ``very massive seed'' in Table~\ref{tab:seed_inversion_ext}, the inverted minima obtained under our  assumptions formally require birth masses \(\gtrsim10^{6}\,M_{\odot}\). This result does not force a unique interpretation: hierarchical mergers of smaller remnants coupled with sustained, near-Eddington accretion can assemble very large black holes within the available cosmic time \citep{Yoo2004,Volonteri2010,Valiante2016}, while super-Eddington (super-critical) accretion episodes, demonstrated in semi-analytic work and multi-dimensional simulations (cosmological semi-analytic models, 3D hydrodynamical circumnuclear simulations and global 3D radiation/magneto-hydrodynamic accretion-flow calculations) to accelerate growth relative to strict Eddington-limited evolution, provide an alternative pathway for stellar-mass seeds to reach high masses rapidly \citep{Lupi2016,Pezzulli2016,Sadowski2016,Inayoshi2020}.

Inspection of Table~\ref{tab:seed_inversion_ext} shows that the sources with the lowest Eddington ratios ($0.04\leq\lambda<0.3$) require ``very massive seeds'', while sources with higher Eddington ratios ($\lambda>0.4$) require ``very light seeds''. That stems from the exponential growth law of Eq.~\ref{eq:Mseed_final} (under the assumption of constant radiative efficiency). Therefore, uncertainties in the Eddington ratio may have a strong impact on inferred seed masses.

This is also exemplified in the specific case of PSO\,J0309+27 that was studied in detail by \cite{Belladitta2022}. 
Using our updated values for PSO\,J0309+27 from Table~\ref{tab:seed_inversion_ext} and replacing the Eddington ratio with the value reported in \cite{Belladitta2022}, namely $\lambda=0.44$, reduces the required seed mass by a factor $\approx 50$. Our formal estimate of $M_{\rm seed}\simeq1.8\times10^{8}\,M_{\odot}$ (based on $\lambda=0.16$) would decrease to $\sim 3.8\times10^{6}\,M_{\odot}$ when adopting their higher Eddington ratio. If we also adopt their final mass estimate ($M_{\rm final}\simeq1.45\times10^{9}\,M_{\odot}$) and their assumed seed formation redshift ($z_0=30$), the inferred initial mass decreases further to $\sim 1.1\times 10^{6}\,M_{\odot}$ under $\eta=0.1$. The difference in the adopted Eddington ratio and formation redshift can largely explain the apparent discrepancy between our seed scenario for PSO\,J0309+27 and the one discussed in \cite{Belladitta2022}.

\section{Conclusions}\label{sec:conclusions}

We have presented a homogeneous analysis of 23 high-redshift blazars spanning from $z\sim 2$ to $z \sim 6$ using archival infrared-to-ultraviolet photometric data and a Bayesian MCMC fitting procedure of the Shakura--Sunyaev accretion disk model. By including the effects of intergalactic medium attenuation, we derived black hole masses and mass accretion rates with quantified uncertainties and avoided the systematic overestimation of $M_{\rm BH}$ (up to a factor of 11) that arises when extinction is neglected. Our results show no evidence for systematic offsets relative to previous literature values, while we recover accretion rates spanning from $\sim$0.05 to $\sim 0.99$ across the sample. We find no evidence of correlation between the black hole masses and Eddington ratios with redshift. We further examined the potential impact of black-hole spin and highlighted how it can influence the derived accretion-disk parameters. With the use of the Eddington-limited growth equation, we placed lower limits on the seed masses required to assemble the observed black holes. The results highlight a broad range of possible seeding channels, from Pop III to direct collapse black holes, with the required seed mass strongly correlated with the used Eddington ratio. 

To our knowledge, this is the first systematic study to model thermal disk emission from a large sample of high-$z$ blazars within a common statistical framework. By providing the largest uniform set of black-hole mass and accretion-rate estimates for these sources with accompanying uncertainties, our results underscore the importance of consistent modeling when comparing across diverse samples. Future progress will rely on improved multiwavelength coverage of the disk emission, as well as independent constraints on spin and accretion histories, which are essential to resolve the degeneracies inherent in SED-based mass determinations.

\section*{Data availability.} Archival data, MCMC fitting results (in .h5 format), and Python scripts to plot results for individual sources are available for download at this link: \url{https://zenodo.org/records/17781963}.

\begin{acknowledgements} 
The authors thank the anonymous referee for their constructive comments that help to improve the manuscript. The authors also thank Andrea Gokus for bringing to their attention another high-$z$ blazar that was added to the analysis. The authors thank Lea Marcotulli for providing the photometric data presented in Marcotulli et al. (2020) and John Kallimanis for useful discussions about statistical methods. M.P. acknowledges support from the Hellenic Foundation for Research and Innovation (H.F.R.I.) under the ``2nd call for H.F.R.I. Research Projects to support Faculty members and Researchers'' through the project UNTRAPHOB (Project ID 3013). G.V. acknowledges support from the H.F.R.I. through the project ASTRAPE (Project ID 7802). 
\end{acknowledgements}

\bibliographystyle{aa} 
\bibliography{refs.bib}

\appendix 
\onecolumn
\section{Representative cases of problematic photometry}
\label{app:phot_examples}

To construct the SEDs we obtained data from the SED builder tool using a 3\arcsec search radius, since the position of the sources is known with high accuracy and can be seen in Table \ref{tab:sample}. In several SEDs we identified, after visual inspection, a couple of outlier points, either yielding too high or too low fluxes. In the case of IR data (e.g., AllWISE), this could be related to the particular field and the large point-spread function of the instrument. For optical data, this could be due to problems in the catalogs used by the SED builder tool. We therefore proceeded to visually inspect the fields to search for possible contamination and also cross-checked the catalog information at the seemingly problematic data points.

\begin{figure}[ht]
    \centering
    \includegraphics[width=0.95\textwidth]{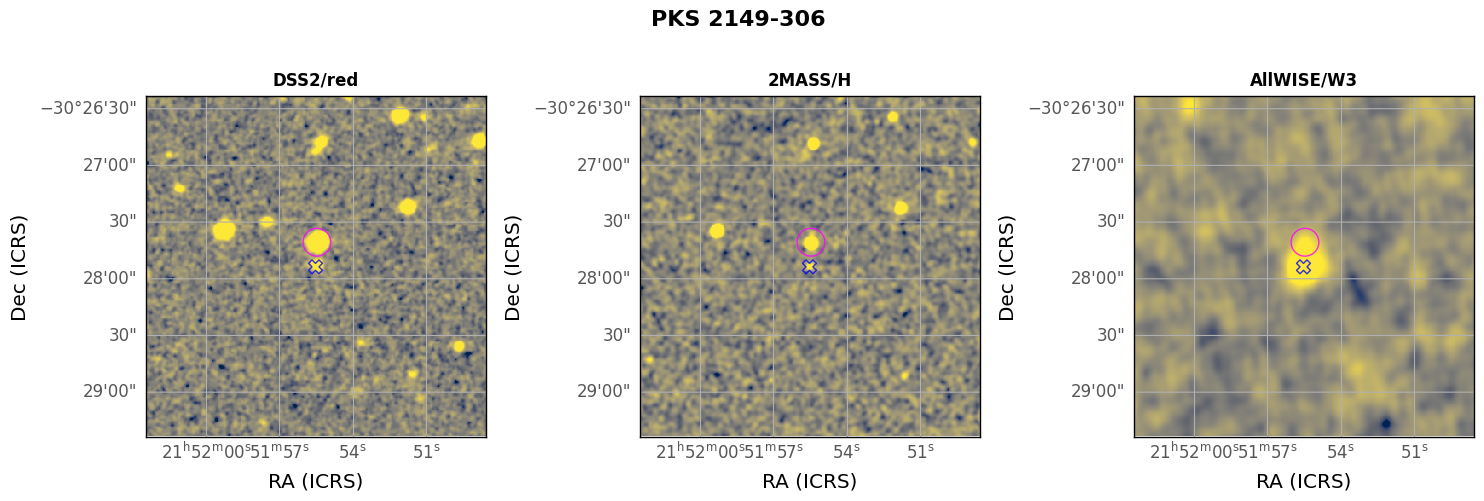}
    \includegraphics[width=0.95\textwidth]{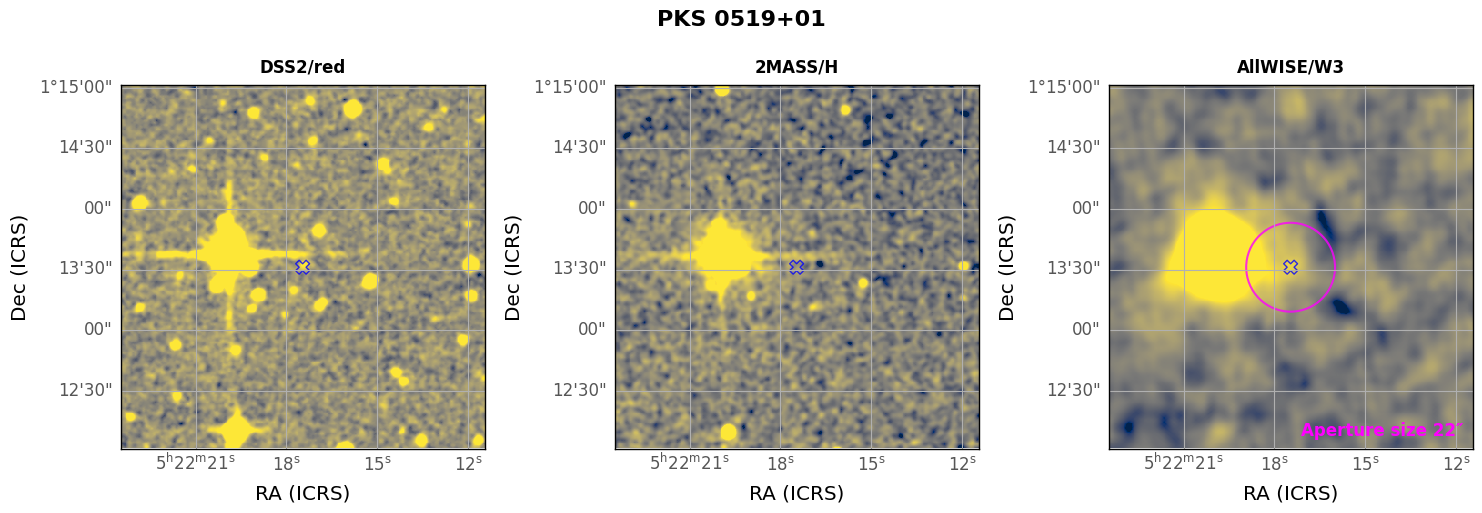}
    \caption{Two representative examples of problematic photometry identified in SED-builder data during our catalog cross-checks and visual inspection. \textit{Top}: The position of PKS\,2149$-$306 is marked with an open x sign, a catalog cross-match with SED builder tool yields a mix of photometric data from the correct counterpart but also from the nearby bright object marked with a magenta circle, which was misidentified as the optical counterpart in the \citet{2009ApJS..184..138H} catalog.  \textit{Bottom}: Same plot for PKS\,0519+01. Contamination from a nearby source is visible in the WISE band when large-aperture photometry is used. The SED Builder tool by default returns fluxes from three different extraction methods without a clear distinction. In both examples as well as in all cases with similar issues, the problematic photometric measurements were excluded from the SEDs used in the analysis.}
    \label{fig:iMAG}    
\end{figure}

A problem we identified in multiple sources was related to the catalog of \citet{2009ApJS..184..138H}. This catalog lists the coordinates of blazars and photometric values (DSS2) for their optical counterparts, without providing the coordinates or an identification for the counterpart. We discovered that the photometric data are not always taken from the correct counterpart, but in some cases from the brightest source located several arc-seconds away. The upper panel in Fig. \ref{fig:iMAG} shows the case of PKS\,2149$-$306, where we indicate the wrong counterpart with a magenta circle.  The three-band image sequence demonstrates that the catalog counterpart does not coincide with the compact optical/IR source identified as the blazar (indicated with a cross). Therefore, we consider this catalog entry to be a misassociation and excluded the corresponding photometry from the SED fitting. 

The lower panel in Fig.~\ref{fig:iMAG} shows the case of PKS\,0519+01, where the fixed-radius $22\arcsec$ aperture provided in the AllWISE catalog \citep{2011ApJ...735..112J} is overplotted as a magenta circle. It is evident that the large fixed aperture encloses flux from a nearby source, producing a biased (overestimated) flux for the blazar when this aperture magnitude is adopted without visual verification. The largest fixed apertures (e.g., the 22\arcsec\ aperture in W3) are susceptible to contamination in crowded fields and for point sources close to neighbors. 

\begin{table*}[h!]
\centering
\caption{Sources with outlier points. The full table, including catalog information from VizieR and NED, is available as supplementary material in the project's Zenodo \href{https://zenodo.org/records/17781963}{repository}.}
\label{tab:sources}
\begin{tabular}{llll}
\hline\hline
Object & Reason for removing outlier points & Catalogs from VizieR & Catalogs from NED \\
\hline
PSO~J0309+27     & Large aperture & GSC 2.4.2, AllWISE, ... &  -- \\
B2~1023+25       & Large aperture + incorrect association & DR14Q UV, DR12Q, ... & B2, SDSS6, ...  \\
SDSS~J0131--0321 & Incorrect association + instrument sensitivity & ATLASRefcat2, DESDR2, ... &  -- \\ 
                 &  limit (noise or contamination) &  &  \\
...              & ... & ... &  ...  \\
\hline
\end{tabular}
\end{table*}

\section{MCMC fitting results}
\label{sec:Results_Fit}
In this section, we provide the tables \ref{tab:fit_params} and \ref{tab:fit_params_power_law} with the MCMC fit results for the sources in our sample. 
To quantitatively compare the SED fits with and without intergalactic medium (IGM) attenuation, we computed the Bayesian Information Criterion (BIC). The resulting BIC values are reported in Table~\ref{tab:fit_params} and provide a measure of the relative goodness of fit while accounting for model complexity.
The attenuation implemented in the model is determined by the source redshift and is kept fixed, hence it does not introduce additional free parameters. Although the unattenuated model occasionally yields lower BIC values, this difference likely reflects statistical fluctuations rather than a genuinely better physical description.
Because attenuation in high-redshift blazars is expected to occur, and our model employs an average effective H I opacity (not adjusted for individual lines of sight), the BIC alone cannot serve as a decisive criterion for model preference. For this reason, the extinction-corrected fits are adopted throughout our figures and analysis, as they offer a physically consistent representation of the source SEDs.

An illustrative example of this limitation is provided by the source B2~1023+25. We were unable to obtain a reliable disk+jet SED fit with attenuation for this object. Inspection of its SED shows that the highest-frequency data points lie systematically above the extinction-corrected model used in our fits (see the supplementary material at the project’s \href{https://zenodo.org/records/17781963}{repository}). Reproducing those points would require a lower optical depth to H\textsc{i} than the effective value we used. Therefore, the true black hole mass and accretion rate values should lie between those obtained with and without extinction (a similar argument can be made also for GB6 B1428+4217).

Another noteworthy case is that of GB6~B1428+4217. \citet{2025ApJ...990..206G} analyzed optical data (in the I and R bands) from this blazar during its 2023–2024 flare and reported a polarization degree (8.6$\pm$0.3)\% in the R band. Such a polarization level indicates a significant contribution from the jet to the observed flux, since the thermal emission from the accretion disk is expected to be largely unpolarized ($<3\%$). In our analysis, we modeled archival data with a pure accretion disk component; however, part of the R-band flux could plausibly originate from a jet component with an as yet unknown polarization. The disk contribution to the observed R- and I-band fluxes can be reduced below the measured values if a black hole mass about ten times larger and an accretion rate roughly three times lower are assumed. This represents an extreme configuration, adopted here only to illustrate the minimal possible contribution of the disk to the total optical emission. At the same time, the available low-frequency data points ($\nu \lesssim 10^{14}$~Hz) are sparse, making it difficult to constrain the jet component directly. This limitation likely explains why a single-component (disk-only) model is statistically preferred in our fits. In such cases, broadband SED modeling of the jet emission, as performed by \citet{2025ApJ...990..206G}, may provide the most reliable means to constrain the disk properties.

\begin{table*}[!htb]
\caption{Fitted model parameters from MCMC analysis without extinction (no ext.) and with extinction (ext.). Uncertainties are statistical and reflect the 68\% range of values derived from the posterior distributions. The objects that have been fitted with the 2 component (jet+disk) model are indicated with a star * by their name. The difference in BIC values from the fits with and without extinction is listed in the last column.}
\label{tab:fit_params}
\centering
\renewcommand{\arraystretch}{1.7}
\setlength{\tabcolsep}{4pt}
\begin{tabular}{l l c cc cc cc cc}
\hline
No. & Name & $z$ &
\multicolumn{2}{c}{$\log_{10}(M_{\mathrm{BH}}/M_{\odot})$} &
\multicolumn{2}{c}{$\log_{10}(\dot{M}/(M_{\odot}\, \rm yr^{-1}))$} &
\multicolumn{2}{c}{$\log_{10}(\lambda_{\mathrm{Edd}})$} & \multicolumn{2}{c}{BIC}\\
\cline{4-5} \cline{6-7} \cline{8-11}  
&  & & no ext. & ext. & no ext. & ext. & no ext. & ext. & no ext. & ext.\\
\hline
1 & PSO J0309+27* & 
6.10 & $9.25_{-0.09}^{+0.08}$ & $9.25_{-0.09}^{+0.07}$ & $0.75^{+0.05}_{-0.04}$ & $0.75^{+0.05}_{-0.04}$  & $-0.9^{+0.1}_{-0.1}$ &   $-0.8^{+0.1}_{-0.1}$ & -20.4 & -20.3 \\  
2 & B2 1023+25* & 5.30 & $9.79^{+0.03}_{-0.04}$ & $9.26^{+0.05}_{-0.04}$ & $0.94^{+0.03}_{-0.03}$ & $1.60^{+0.05}_{-0.04}$ &  $-1.20^{+0.02}_{-0.02}$ & $-0.007^{+0.005}_{-0.013}$ & 97.3 & 410.2 \\
3 & SDSS J0131-0321* & 5.18 & $10.34^{+0.03}_{-0.03}$ & $9.72^{+0.05}_{-0.08}$ & $1.70^{+0.03}_{-0.03}$ & $1.76^{+0.02}_{-0.02}$ & $-0.98^{+0.02}_{-0.02}$ & $-0.30^{+0.07}_{-0.05}$ & 168.3 & -26.1  \\ 
4 & GB6 B1428+4217 & 4.715 & $9.93^{+0.05}_{-0.05}$ & $8.90^{+0.11}_{-0.06}$ & $1.09^{+0.05}_{-0.05}$ & $1.09^{+0.06}_{-0.09}$ & $-1.19^{+0.02}_{-0.02}$ & $>-0.3$ & 36.7  & 113.9\\
5 & PMN J2134-0419* & 4.35 & $9.73^{+0.04}_{-0.04}$ &  $9.13^{+0.06}_{-0.07}$ &  $1.00^{+0.03}_{-0.03}$ & $1.15^{+0.03}_{-0.03}$ & $-1.07^{+0.03}_{-0.03}$ & $-0.32^{+0.09}_{-0.08}$ & 44.1 & -38.4 \\
6 & J151002+570243* & 4.31 & $9.85^{+0.03}_{-0.03}$ & $8.8^{+0.1}_{-0.1}$ & $0.91^{+0.03}_{-0.03}$ & $0.95^{+0.04}_{-0.04}$ & $-1.28^{+0.01}_{-0.01}$ & $-0.20^{+0.1}_{-0.1}$ & -39.1 & -11.3\\
7 & J222032.50+002537.5* & 4.22 & $9.65^{+0.02}_{-0.02}$ & $9.08^{+0.06}_{-0.06}$ & $0.76^{+0.01}_{-0.01}$ & $0.83^{+0.02}_{-0.02}$ & $-1.24^{+0.01}_{-0.01}$ & $-0.59^{+0.07}_{-0.06}$ & -246.0 & -16.6\\ 
8 & J142048.01+120545.9* & 4.03 & $9.46^{+0.02}_{-0.02}$ & $8.65^{+0.02}_{-0.02}$ & $0.76^{+0.01}_{-0.01}$ & $0.97^{+0.01}_{-0.01}$ & $-1.04^{+0.01}_{-0.01}$ & $-0.02^{+0.01}_{-0.03}$ & -195.7 & -182.3 \\
9 & J135406-020603 & 3.71 & $9.80^{+0.10}_{-0.09}$ & $9.2^{+0.1}_{-0.1}$ & $0.96^{+0.07}_{-0.07}$ & $1.06^{+0.05}_{-0.05}$ & $-1.19^{+0.08}_{-0.07}$ & $-0.5^{+0.2}_{-0.1}$ & -40.9 & 13.13 \\
10 & J020346+113445 & 3.63 & $9.80^{+0.10}_{-0.09}$ & $9.59^{+0.07}_{-0.07}$ & $0.96^{+0.07}_{-0.07}$ & $0.99^{+0.03}_{-0.03}$ & $-1.19^{+0.08}_{-0.07}$ & $-0.95^{+0.07}_{-0.07}$ & 12.8 & -10.0 \\ 
11 & J013126-100931 & 3.51 & $9.66^{+0.06}_{-0.06}$ & $9.46^{+0.05}_{-0.05}$ & $0.51^{+0.05}_{-0.04}$ & $0.47^{+0.03}_{-0.03}$ & $-1.50^{+0.03}_{-0.03}$ & $-1.33^{+0.03}_{-0.03}$ & -6.9 & -40.6 \\
12 & J064632+445116 & 3.41 & $9.88^{+0.04}_{-0.04}$ & $9.65^{+0.05}_{-0.05}$ & $1.29^{+0.03}_{-0.03}$ & $1.33^{+0.02}_{-0.02}$ & $-0.94^{+0.03}_{-0.03}$ & $-0.66^{+0.06}_{-0.06}$ & 0.7 & -23.4\\
13 & S5 0014+813* & 3.366 & $10.23^{+0.03}_{-0.04}$ & $9.95^{+0.02}_{-0.02}$ & $2.16^{+0.03}_{-0.03}$ & $2.29^{+0.02}_{-0.02}$ & $-0.41^{+0.02}_{-0.02}$ & $>-0.033$ & -12.4 & -38.8\\
14 & J212912-153841 & 3.26 & $10.25^{+0.05}_{-0.05}$ & $9.95^{+0.03}_{-0.03}$ & $1.89^{+0.04}_{-0.04}$ & $1.93^{+0.01}_{-0.01}$ & $-0.71^{+0.03}_{-0.03}$ & $-0.36^{+0.02}_{-0.02}$ & 16.5 & -74.1 \\ 
15 & PKS 0537-286* & 3.104 & $9.70^{+0.04}_{-0.04}$ & $9.29^{+0.06}_{-0.08}$ & $0.95^{+0.03}_{-0.03}$ & $0.92^{+0.02}_{-0.03}$ & $-1.10^{+0.03}_{-0.02}$ & $-0.71^{+0.07}_{-0.06}$ & -77.4 & -134.1 \\
16 & GB6 0805+614 & 3.033 & $9.79^{+0.04}_{-0.04}$ & $9.76^{+0.04}_{-0.04}$ & $0.42^{+0.09}_{-0.15}$ & $0.74^{+0.03}_{-0.02}$ & $-1.40^{+0.04}_{-0.09}$ & $-1.12^{+0.19}_{-0.08}$ & -40.0 & -22.8\\
17 & PKS 0519+01* & 2.94 & $9.47^{+0.07}_{-0.10}$ & $9.1^{+0.2}_{-0.3}$ & $0.42^{+0.09}_{-0.15}$ & $0.3^{+0.2}_{-0.2}$ & $-1.39^{+0.04}_{-0.09}$ & $-1.12^{+0.20}_{-0.08}$ & -117.2 & -122.5  \\
18 & PMN J1344-1723* & 2.49 & $8.4^{+0.3}_{-0.1}$ & $8.4^{+0.2}_{-0.1}$ & $0.49^{+0.08}_{-0.10}$ & $0.50^{+0.07}_{-0.10}$ & $-0.3^{+0.2}_{-0.4}$ & $>-0.5$ & -117.3 & -65.7 \\
19 & TXS 1149-084* & 2.37 & $8.52^{+0.11}_{-0.07}$ & $8.52^{+0.10}_{-0.06}$ & 
$0.72^{+0.04}_{-0.06}$ & $0.74^{+0.04}_{-0.05}$ & $-0.1^{+0.1}_{-0.2}$ & $>-0.2$ & -124.1 & -119.2 \\
20 & PKS 2149-306* & 2.345 & $9.77^{+0.02}_{-0.02}$ & $9.55^{+0.02}_{-0.02}$ &
$1.20^{+0.01}_{-0.01}$ & $1.21^{+0.01}_{-0.01}$ & $-0.91^{+0.01}_{-0.01}$ & $-0.68^{+0.02}_{-0.02}$  & -185.2 & -281.8 \\
21 & 4C 71.07 (0836+710)* & 2.172 &
$9.79^{+0.01}_{-0.01}$ & $9.59^{+0.02}_{-0.02}$ & $1.494^{+0.008}_{-0.009}$ & $1.543^{+0.009}_{-0.009}$ & 
$-0.650^{+0.009}_{-0.009}$ & $-0.40^{+0.01}_{-0.01}$ & -397.0 &  -396.8 \\ 
22 & PKS 0227-369* & 2.12 & $8.90^{+0.06}_{-0.07}$ &  $8.5^{+0.1}_{-0.1}$ &  $0.36^{+0.03}_{-0.03}$ & $0.56^{+0.06}_{-0.05}$ & $-0.89^{+0.08}_{-0.07}$ & $-0.3^{+0.2}_{-0.2}$ & -67.2 & -72.2 \\
23 & PKS 0528+134* & 2.04 & $9.3^{+0.1}_{-0.1}$ & $9.29^{+0.11}_{-0.09}$ & $1.43^{+0.07}_{-0.11}$ & $1.46^{+0.07}_{-0.06}$ & $-0.2^{+0.1}_{-0.2}$ & $-0.2^{+0.1}_{-0.2}$ & 10.2& 8.7 \\
\hline
\end{tabular}
\end{table*}

\begin{table*}[!htb]
\caption{Best-fit parameters of the power-law component (with and without extinction) of the jet+disk model.}
\label{tab:fit_params_power_law}
\centering
\renewcommand{\arraystretch}{1.7}
\setlength{\tabcolsep}{8pt}
\begin{tabular}{r l cc cc cc}
\hline
No. & Name &
\multicolumn{2}{c}{$\log_{10}F_0$ (cgs units)} &
\multicolumn{2}{c}{$\log_{10}\nu_c$ (Hz)} &
\multicolumn{2}{c}{$s$} \\
 & & no ext. & ext. & no ext. & ext. & no ext. & ext. \\
\hline
1 & PSO J0309+27 & $-11.9^{+0.1}_{-0.1}$ & $-11.9^{+0.1}_{-0.1}$ & $13.9^{+0.2}_{-0.1}$ & $13.9^{+0.2}_{-0.1}$ & $-2.6^{+0.3}_{-0.2}$ & $-2.6^{+0.3}_{-0.3}$ \\
2 & B2 1023+25 & $-11.6^{+0.3}_{-0.4}$ & $-11.8^{+0.5}_{-0.6}$ & $13.7^{+0.2}_{-0.1}$ & $13.8^{+0.3}_{-0.2}$ & $-1.29^{+0.07}_{-0.07}$ & $-1.000^{+0.500}_{-0.004}$ \\
3 & SDSS J0131$-$0321 & $-12.4^{+0.2}_{-0.2}$ & $-12.5^{+0.1}_{-0.1}$ & $13.9^{+0.2}_{-0.1}$ & $14.2^{+0.2}_{-0.2}$ & $-1.13^{+0.05}_{-0.05}$ & $-1.63^{+0.06}_{-0.06}$ \\ 
5 & PMN J2134$-$0419 & $-11.6^{+0.3}_{-0.4}$ & $-11.7^{+0.3}_{-0.3}$ & $13.7^{+0.2}_{-0.1}$ & $13.8^{+0.2}_{-0.2}$ & $-1.48^{+0.08}_{-0.09}$ & $-1.72^{+0.08}_{-0.07}$ \\
6 & J151002+570243 & $-13.0^{+0.7}_{-0.4}$ & $-13.0^{+0.3}_{-0.2}$ & $13.8^{+0.5}_{-0.2}$ & $14.9^{+0.4}_{-0.4}$ & $-1.80^{+0.08}_{-0.08}$ & $-1.66^{+0.09}_{-0.09}$ \\
7 & SDSS J2220+0025 & $-11.5^{+0.3}_{-0.3}$ & $-12.5^{+0.1}_{-0.2}$ & $13.6^{+0.1}_{-0.1}$ & $14.8^{+0.4}_{-0.5}$ & $-1.98^{+0.04}_{-0.04}$ & $-1.62^{+0.04}_{-0.05}$ \\
8 & SDSS J1420+1205 & $-12.1^{+0.2}_{-0.2}$ & $-12.8^{+0.1}_{-0.1}$ & $13.8^{+0.2}_{-0.1}$ & $15.3^{+0.1}_{-0.2}$ & $-2.06^{+0.05}_{-0.05}$ & $-1.98^{+0.05}_{-0.05}$ \\
13 & S5 0014+813 & $-11.7^{+0.4}_{-0.4}$ & $-11.6^{+0.3}_{-0.2}$ & $14.1^{+0.8}_{-0.4}$ & $14.4^{+0.7}_{-0.5}$ & $-1.7^{+0.1}_{-0.1}$ & $-2.3^{+0.1}_{-0.1}$    \\
15 & PKS 0537-286 & $-12.7^{+0.2}_{-0.2}$ & $-12.7^{+0.1}_{-0.1}$ & $14.6^{+0.5}_{-0.2}$ & $14.6^{+0.3}_{-0.1}$ & $-1.82^{+0.07}_{-0.06}$ & $-2.03^{+0.07}_{-0.06}$    \\
17 & PKS 0519+01 &$-12.8^{+0.1}_{-0.1}$ & $-12.8^{+0.1}_{-0.1}$ & $14.9^{+0.2}_{-0.3}$ & $14.9^{+0.2}_{-0.2}$ & $-2.19^{+0.08}_{-0.08}$ & $-2.21^{+0.08}_{-0.08}$   \\ 
18 & PMN J1344$-$1723 & $-11.70^{+0.07}_{-0.07}$ &$-11.70^{+0.06}_{-0.05}$ & $15.45^{+0.04}_{-0.08}$ & $14.47^{+0.10}_{-0.03}$ & $-1.7^{+0.1}_{-0.1}$ & $-1.9^{+0.2}_{-0.1}$ \\
19 & TXS 1149$-$084 & $-11.86^{+0.05}_{-0.05}$ & $-11.86^{+0.05}_{-0.05}$ & $14.7^{+0.2}_{-0.1}$ & $14.7^{+0.1}_{-0.1}$ & $-2.07^{+0.07}_{-0.07}$ & $-2.04^{+0.08}_{-0.07}$ \\
20 & PKS 2149$-$306 & $-11.9^{+0.1}_{-0.1}$ & $-11.98^{+0.06}_{-0.05}$ & $14.2^{+0.4}_{-0.2}$ & $14.7^{+0.5}_{-0.3}$ & $-1.79^{+0.04}_{-0.04}$ & $-1.96^{+0.05}_{-0.04}$ \\
21 & 4C 71.07 (0836+710) & $-11.5^{+0.1}_{-0.1}$ & $-11.62^{+0.05}_{-0.05}$ & $14.3^{+0.7}_{-0.4}$ & $15.2^{+0.2}_{-0.3}$ & $-2.13^{+0.05}_{-0.04}$ & $-2.14^{+0.05}_{-0.04}$ \\
22 & PKS 0227$-$369 & $-12.1^{+0.1}_{-0.1}$ & $-12.11^{+0.08}_{-0.08}$ & $14.3^{+0.3}_{-0.1}$ & $14.5^{+0.4}_{-0.3}$ & $-1.90^{+0.08}_{-0.08}$ & $-1.9^{+0.1}_{-0.1}$ \\
23 & PKS 0528+134 & $-12.1^{+0.4}_{-0.3}$ & $-12.1^{+0.4}_{-0.3}$ & $14.2^{+0.8}_{-0.5}$ & $14.2^{+0.8}_{-0.5}$ & $-1.7^{+0.3}_{-0.2}$ & $-1.8^{+0.3}_{-0.2}$ \\
\hline
\end{tabular}
\end{table*}

\begin{figure*}[htbp!]
    \centering
    \includegraphics[width=0.33\textwidth]{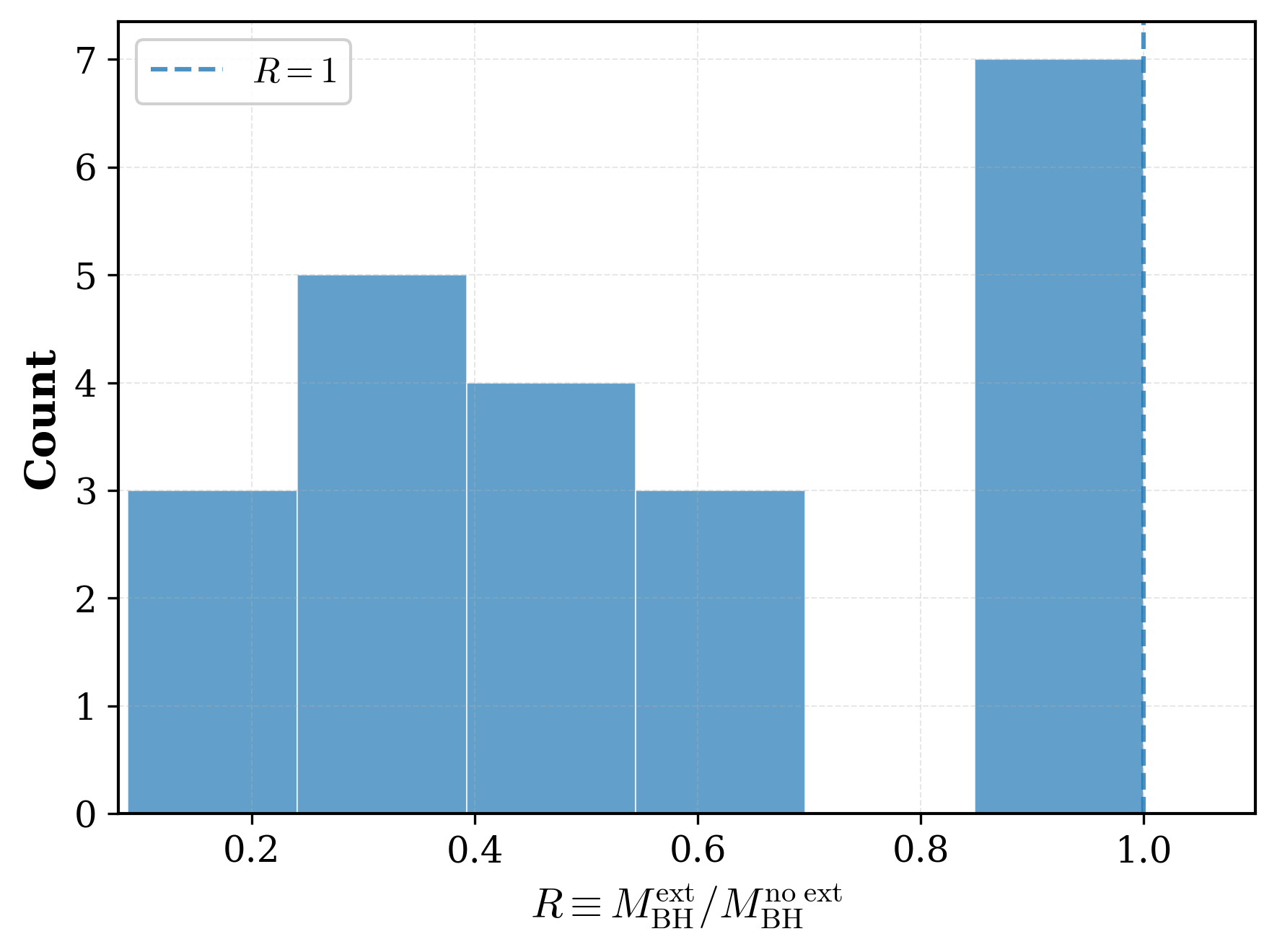}
    \includegraphics[width=0.33\textwidth]{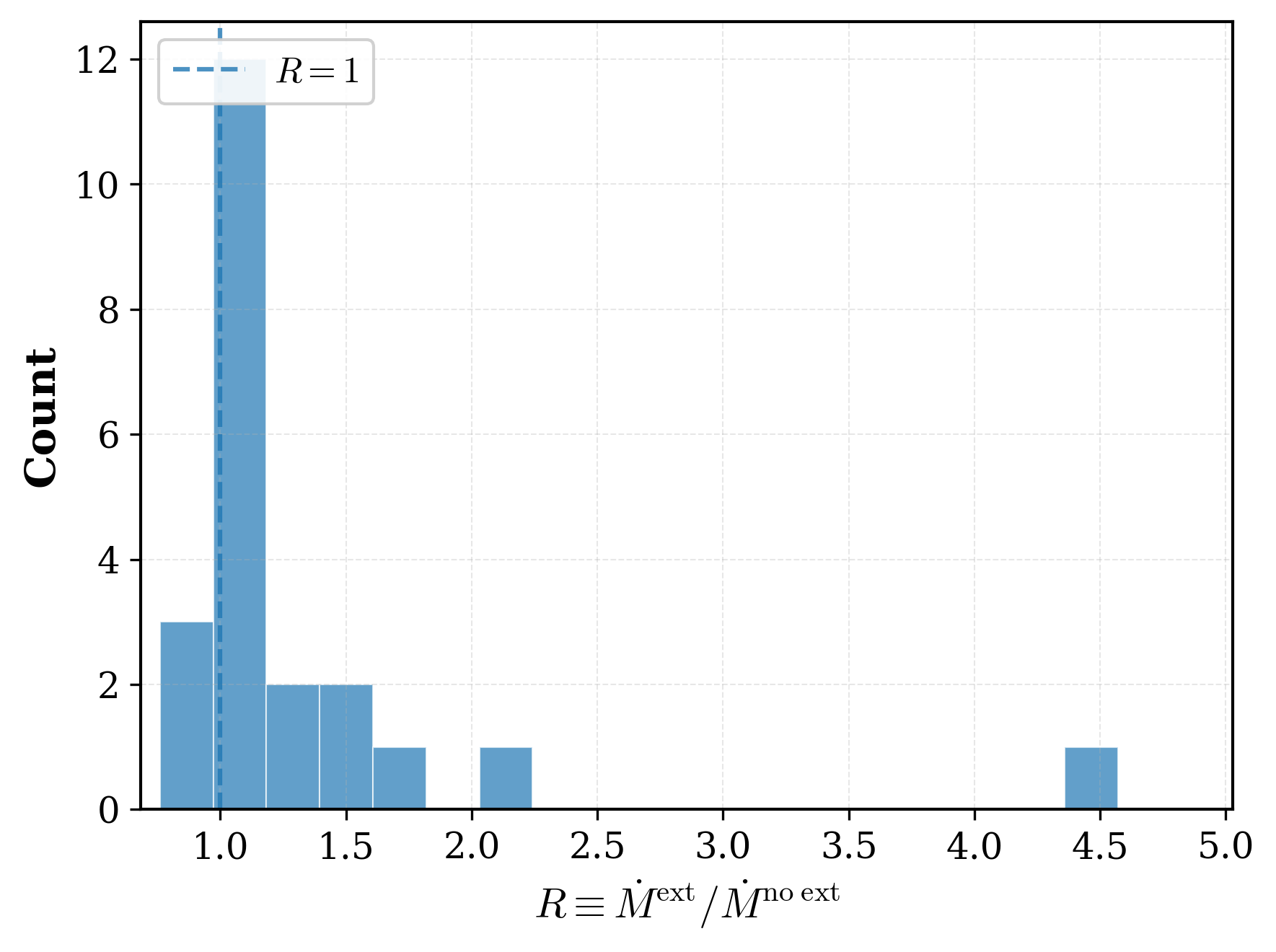}
    \includegraphics[width=0.33\textwidth]{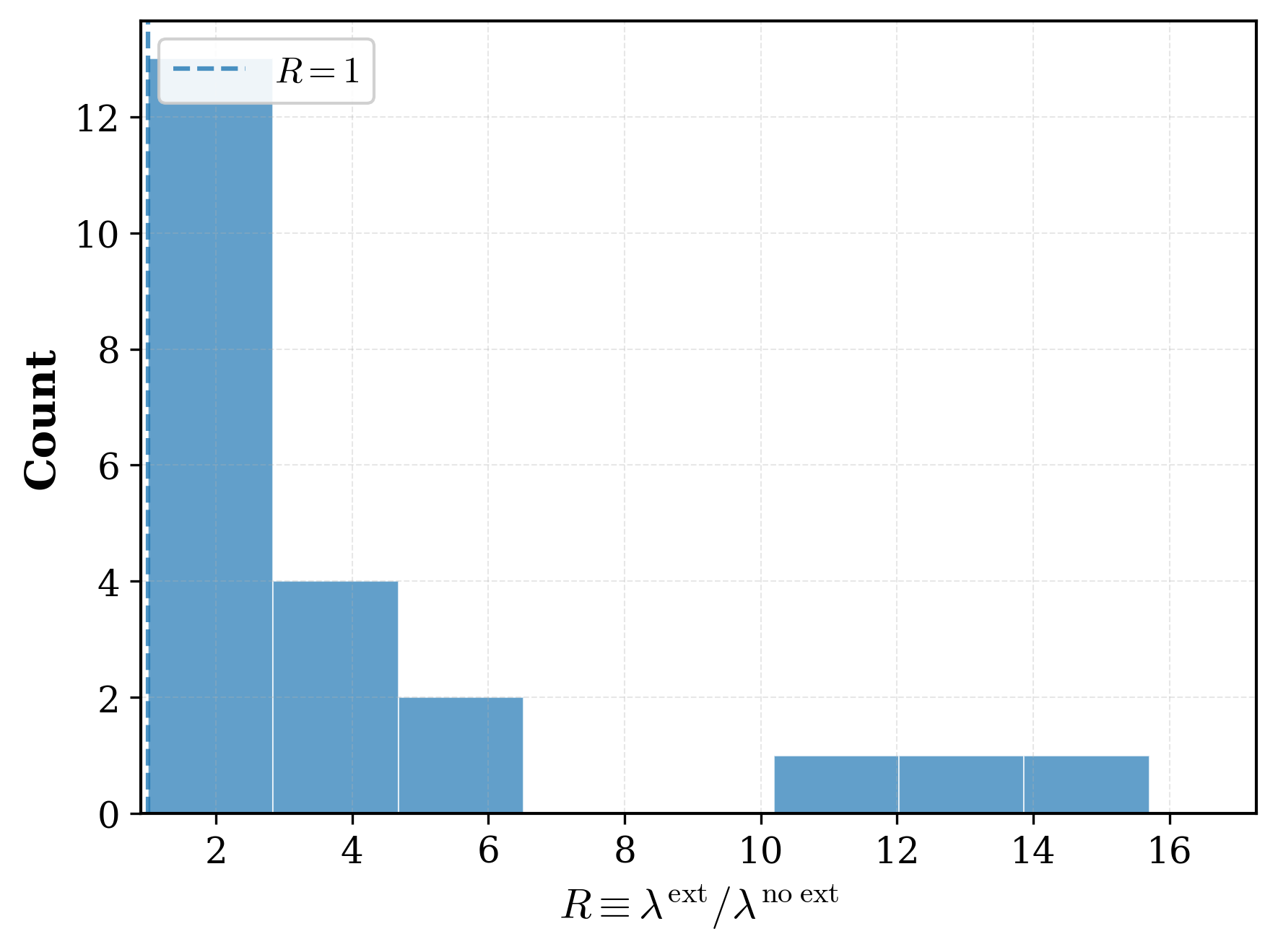}
    \caption{
        Ratios of quantities derived with and without accounting for extinction. 
        The top row shows the ratio of black hole masses (left) 
        and the ratio of accretion rates (right), 
        while the bottom panel shows the ratio of Eddington ratios. 
        The dashed black vertical line at $R=1$ corresponds to no change between the two fits .
        These plots highlight the systematic trend: 
        neglecting extinction leads to overestimates of black hole masses 
        and underestimates of Eddington ratios. The blazars B2~1023+25 and GB6 B1428+4217 are excluded from this figure because of reasons explained in Appendix \ref{sec:Results_Fit}.}
    \label{fig:ratios_appendix}
\end{figure*}

\section{Impact of extinction on derived quantities}\label{appC}

In this section, we quantify how the inclusion of IGM attenuation affects the disk-derived parameters by directly comparing posterior medians obtained with and without absorption. The comparison shown in Fig.~\ref{fig:ratios_appendix} demonstrates how neglecting this effect leads to systematically higher black hole masses and lower Eddington ratios. These results highlight that IGM attenuation must be accounted for in order to obtain more reliable and accurate estimates of the derived quantities.

\section{Black hole mass - accretion rate: empirical modeling with spin-induced covariance}
\label{appD}

We model the relationship between black hole mass $M_{\rm BH}$, and accretion rate $\dot M$, in logarithmic space using a linear relation
\begin{equation}
\log_{10}\!\left(\frac{M_{{\rm BH}}}{10^{9}\,M_\odot}\right)
\;=\; b \;+\; m\,\log_{10}\!\left(\frac{\dot M}{1\,M_\odot\,{\rm yr}^{-1}}\right).
\label{eq:mbh-mdot}
\end{equation}
where $m$ is the slope and $b$ the offset. 
The masses and accretion rates used in the fit were taken from our spectral MCMC runs performed under the assumption of non-rotating black holes (\(a=0\)). Because the true spins are unknown, this induces a systematic degeneracy between the inferred black-hole mass \(M_{\rm BH}\) and the accretion rate \(\dot{M}\) via the spin-dependent inner radius \(R_{\rm ISCO}(a)\) (Eq.~\ref{eq:scale}).
This spin-induced uncertainty produces an additive shift $\delta = \log_{10}(R_{\rm in,ref}/R_{\rm in})$ in $\log_{10} M_{\rm BH}$ and $-\delta$ in $\log_{10} \dot M$, introducing a covariance along the $y=-x$ direction. 

This effect can be represented as a covariance matrix per-object in logarithmic space, 
\begin{equation}
\mathbf{\Sigma}_{\delta} = \mathrm{Var}(\delta) 
\begin{pmatrix}
1 & -1 \\
-1 & 1
\end{pmatrix}
\label{eq:spin_mat}
\end{equation}
which is added to the measurement covariance, effectively capturing the scatter induced by the spin distribution. The total residual perpendicular to the linear relation, $d_\perp$, is assumed to be Gaussian distributed. The corresponding log-likelihood is then
\begin{equation}
\ln \mathcal{L} = \sum_i -\frac{1}{2} \left[ \ln \left(2\pi \, \sigma_{\perp,i}^2 \right) + \frac{d_{\perp,i}^2}{\sigma_{\perp,i}^2} \right],
\end{equation}
where $\sigma_{\perp,i}^2 = \mathbf{n}^\mathrm{T} (\mathbf{\Sigma}_{{\rm meas},i} + \mathbf{\Sigma}_{\delta,i}) \mathbf{n} + \sigma_\perp^2$, $\mathbf{n}$ is the unit normal vector to the line, and $\sigma_\perp$ accounts for additional intrinsic scatter.

To account for measurement errors, each source is represented by a ``cloud'' of \(K\) realizations in logarithmic space:
\[
x_{i,j}\sim\mathcal{N}(x_i,\sigma_{x,i}^2),\qquad
y_{i,j}\sim\mathcal{N}(y_i,\sigma_{y,i}^2),\qquad j=1,\ldots,K,
\]
where \(x_i=\log_{10}\dot{M}_i\) (with \(\dot{M}\) in \(M_\odot\ \mathrm{yr}^{-1}\)) and \(y_i=\log_{10}M_{{\rm BH},i}\) (with \(M_{\rm BH}\) expressed in units of \(10^{9}\,M_\odot\)). For each draw, we sample a spin \(a_{i,j}\sim\mathcal{U}[0,1)\), compute \(R_{\rm ISCO}(a_{i,j})\) from Eq.~\ref{eq:Kerr}, rescale the linear-space pair \(\{M_{\rm BH},\dot{M}\}\) according to Eq.~\ref{eq:scale} with reference radius \(R_{\rm ref}=6\,R_g\), and then map the rescaled values back to \((x_{i,j},y_{i,j})\).

\begin{figure}[ht]
    \centering
    \includegraphics[width=0.85\textwidth]{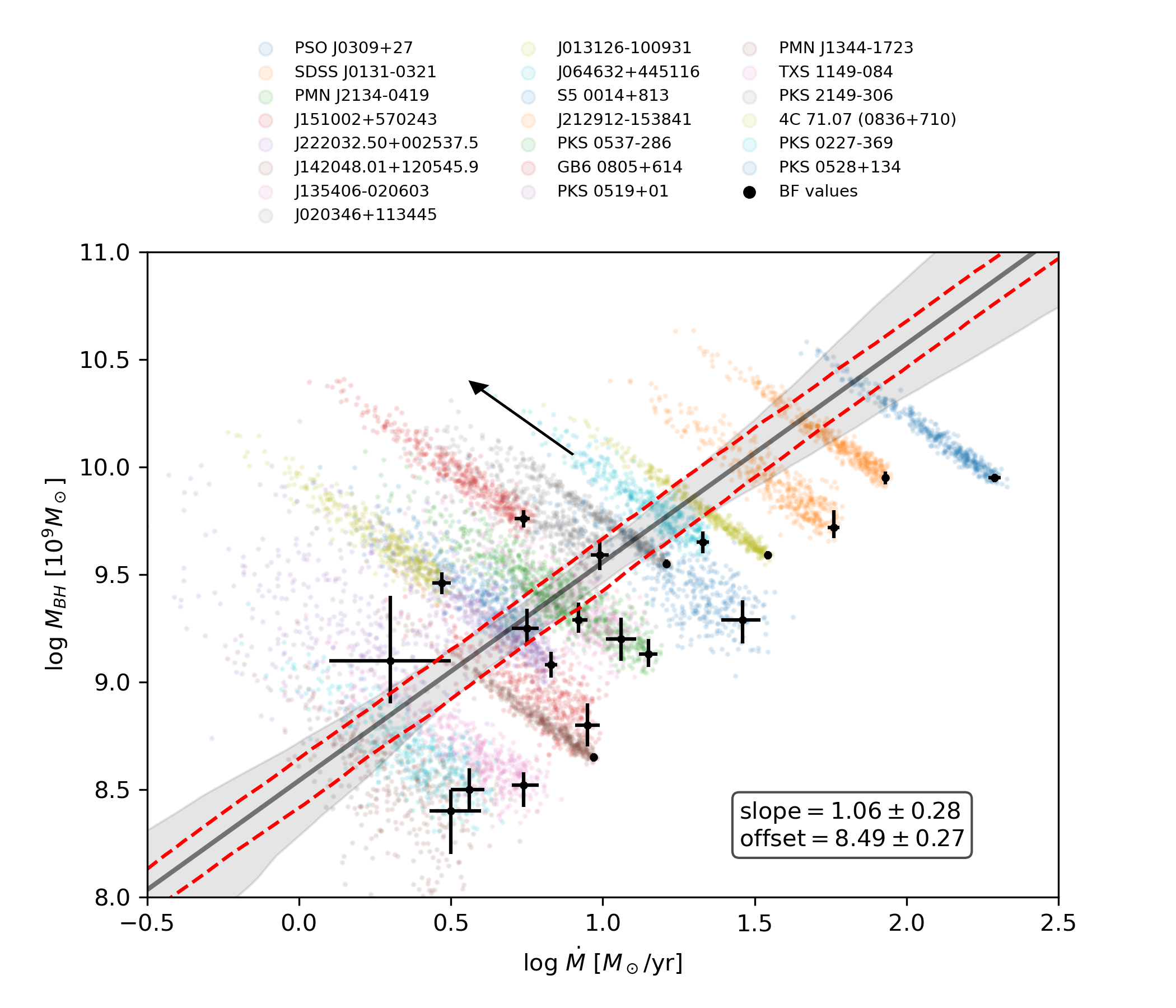}
    \caption{Log-log plot of $M_{\rm BH}$ and $\dot M$ plot based on our fit results assuming zero BH spin (black points indicate the median values from the posterior distributions). Colored clouds are computed based on statistical uncertainties of the fit, as well as a correction for uniform spin following Eq.~\ref{eq:spin_mat} that causes a scatter in the direction of the arrow (see text for details). The population is fitted by a linear model defined in Eq.~\ref{eq:mbh-mdot}. The median and the 68\% credible region are marked by the gray line and the shaded region respectively. Red lines mark the median excess scatter predicted by the fit due to BH spin uncertainty.}
    \label{fig:fit}
\end{figure}

Finally, we sample the posterior distribution of $(m,b,\sigma_\perp)$ using \texttt{UltraNest}, marginalizing over both spin-induced uncertainty and measurement errors. The resulting posterior provides the best-fit linear relation and associated credible intervals. In Fig.~\ref{fig:fit} we plot the fitted $M_{\rm BH}$ and $\dot{M}$ values with their errors, together with the clouds of realizations based on the spin distributions. We also indicate the direction of the spin-induced scatter with an arrow. The best-fit linear model is shown with a gray shaded area and solid line, while the dashed red lines represent the fitted scatter. 

\end{document}